\documentclass[10pt,twocolumn,twoside]{IEEEtran}
\usepackage{amsmath,amsfonts}
\usepackage{algorithm}
\usepackage{algpseudocode}
\usepackage{array}
\usepackage[caption=false,font=normalsize,labelfont=sf,textfont=sf]{subfig}
\usepackage{textcomp}
\usepackage{stfloats}
\usepackage{url}
\usepackage{verbatim}
\usepackage{graphicx}
\usepackage{cite}
\usepackage{amsfonts}       
\usepackage{nicefrac}       
\usepackage{microtype}      
\usepackage{xcolor}         
\usepackage{graphicx}
\usepackage{caption}
\usepackage{booktabs}
\usepackage{amsmath,amsthm}
\usepackage{cite}
\usepackage[colorlinks,bookmarks=false]{hyperref}
\hypersetup{citecolor=[RGB]{0,0,255}, urlcolor=[RGB]{0,0,255}}
\DeclareMathOperator*{\argmin}{argmin}
\newtheorem{proposition}{Proposition}
\newtheorem{definition}{Definition}
\newtheorem{theorem}{Theorem}
\newtheorem{corollary}{Corollary}
\newtheorem{assumption}{Assumption}
\newtheorem{remark}{Remark}

\newcommand{\revise}[1]{{\color{black} #1}}

\newcommand{\dist}{\mathrm{dist}}

\begin{document}
\renewcommand{\algorithmicrequire}{\textbf{Input:}}
\renewcommand{\algorithmicensure}{\textbf{Initialization:}}
\def\NoNumber#1{{\def\alglinenumber##1{}\State #1}\addtocounter{ALG@line}{-1}}
\title{Stability Constrained Reinforcement Learning \\for Decentralized Real-Time Voltage Control}

\author{Jie Feng~\IEEEmembership{Student Member,~IEEE,} Yuanyuan Shi~\IEEEmembership{Member,~IEEE,} Guannan Qu~\IEEEmembership{Member,~IEEE,} Steven H. Low~\IEEEmembership{Fellow,~IEEE,} Anima Anandkumar~\IEEEmembership{Fellow,~IEEE,} and Adam Wierman~\IEEEmembership{Member,~IEEE}
\thanks{The authors are supported by NSF grant ECCS-2200692.}
\thanks{Jie Feng and Yuanyuan Shi are with the Department of Electrical and Computer Engineering, University of California San Diego {\tt\small jif005,yus047@ucsd.edu}.}
\thanks{Guannan Qu is with the Department of Electrical and Computer Engineering, Carnegie Mellon University.}
\thanks{Steven H. Low, Anima Anandkumar, and Adam Wierman are with the Computing and Mathematical Sciences Department, Caltech.}}

\maketitle

\begin{abstract}
Deep reinforcement learning has been recognized as a promising tool to address the challenges in real-time control of power systems. However, its deployment in real-world power systems has been hindered by a lack of explicit stability and safety guarantees. In this paper, we propose a stability-constrained reinforcement learning (RL) method for real-time implementation of voltage control, that guarantees system stability both during policy learning and deployment of the learned policy. The key idea underlying our approach is an explicitly constructed Lyapunov function that leads to a sufficient structural condition for stabilizing policies, i.e., monotonically decreasing policies guarantee stability. We incorporate this structural constraint with RL, by parameterizing each local voltage controller using a monotone neural network, thus ensuring the stability constraint is satisfied by design.
We demonstrate the effectiveness of our approach in both single-phase and three-phase IEEE test feeders, where the proposed method can reduce the transient control cost by more than 26.7\% and shorten the voltage recovery time by 23.6\% on average compared to the widely used linear policy, while always achieving voltage stability. In contrast, standard RL methods often fail to achieve voltage stability. 
\end{abstract}


\section{Introduction}
\label{sec:intro}
The voltage control problem is one of the most critical problems in the control of power network. The primary purpose of voltage control is to maintain the voltage magnitude within an acceptable range under all possible working conditions. Due to the recent proliferation of distributed energy resources (DERs) such as solar and electric vehicles, voltage deviations are becoming increasingly complex and unpredictable. As a result, conventional voltage regulation methods based on on-load tap changing transformers, capacitor banks, and voltage regulators\cite{Baran1989a,Turitsyn11} may fail to respond to the rapid and possibly large fluctuations. 
\revise{
DERs can adjust the reactive power output based on the real-time voltage measurement to achieve fast and flexible voltage stabilization~\cite{8332112}.}

To coordinate the inverter-connected resources for real-time voltage control, one key challenge is to design control rules that can stabilize the system at scale with \emph{limited information}. 
Despite the progress, most of the existing work has only been able to optimize the steady-state cost, i.e. the cost of the operating point after the voltage converges (see~\cite{Farivar-2012-VVC-PES,zhu2016fast,qu2019optimal,chen_input_2020} and references within). \revise{However, as the system is subject to more frequent load and generation fluctuations, optimizing the transient performance becomes of equal importance.} Once a voltage violation happens, an important goal is to bring the voltage profile back to the safety region as soon as possible, at the minimum control costs. 

Optimizing or even analyzing the transient cost of voltage control has long been challenging as this is a nonlinear control problem \cite{shi2021stability}. The challenge is further complicated by the fact that exact model of a distribution system is often unknown due to frequent system reconfigurations~\cite{weng2016distributed} and limited communication infrastructure.
Reinforcement Learning (RL) has emerged as a promising tool to address this challenge. 
One intriguing benefit of RL methods is their model-free characteristic, which means no prior knowledge of the system models is required. Further, with the introduction of neural networks to RL, deep reinforcement learning has great expressive power and has shown impressive results in learning nonlinear controllers with good transient performance.

Despite the promising attempts, one difficulty in applying RL for voltage control is the lack of stability guarantee~\cite{regulator,haes2019survey}. 
Even if the learned policy may appear ``stable'' on a training data set, it is not guaranteed to be stable in unseen test cases and stability requires explicit characterization. 
Motivated by this challenge, the question we address in this paper is: 
\begin{center}
    \emph{Can RL be applied for voltage control with a provable stability guarantee?} 
\end{center}

The key idea underlying our approach is that, with a judiciously chosen Lyapunov function, 
strict \emph{monotonicity} of the policy is sufficient to guarantee voltage stability (Theorem \ref{thm:voltage_stab}). 
\revise{Given that monotonicity is a \emph{model-free} constraint, it is practical to design a stabilizing RL controller without model knowledge.} 
To enforce this structural constraint, we propose a decentralized controller (Stable-DDPG, Algorithm \ref{alg:stable_ddpg}) which integrates the stability constraint with a popular RL framework deep deterministic policy gradient (DDPG) \cite{lillicrap2015continuous} through monotone policy network design. The proposed method enables us to leverage the power of deep RL to improve the transient performance of voltage control without knowing the underlying model parameters. 
We conduct extensive numerical case studies in both single-phase and three-phase IEEE test feeders to demonstrate the effectiveness of the proposed Stable-DDPG with both simulated voltage disturbances and real-world data. \revise{The trained Stable-DDPG can compute control actions efficiently (within 1 ms), which facilitates real-time implementation of neural network-based voltage controllers.} This paper extends the result of our previous conference version \cite{shi2021stability} in the following aspects:
\begin{itemize}
    \item We extend the stability analysis from continuous-time to discrete-time systems to better accommodate the discrete-time nature of inverter-based controllers.
    \item We construct a new discrete-time Lyapunov function and derive the structural constraints for stabilizing controller in Theorem \ref{thm:voltage_stab}. The discrete-time stability constraint requires the policy to be monotonically decreasing, and lower bounded by a value related to the sampling time. The clear relationship between stability and sampling time can assist the practical implementation of the proposed voltage controller with a finite sampling time. As the sampling time $\Delta T \rightarrow 0$, the stability condition reduces to the continuous-time stability condition. 
    \item We test the proposed approach through extensive numerical studies in IEEE single-phase and three-phase systems with simulated and real-world data.
\end{itemize}

\subsection{Related work}
\paragraph{Steady-state cost optimization} Existing literature in optimizing the steady-state cost of voltage control can be roughly classified into two categories based on \cite{qu2019optimal}: feed-forward optimization methods and feedback control methods. 
A typical example of feed-forward optimization methods is Optimal Power Flow (OPF) based methods\cite{Farivar-2012-VVC-PES,7857812,6963439,7042735,9173521}, where control actions are calculated by solving an optimization problem to minimize the power loss subject to voltage constraints. 
\revise{These algorithms assume knowledge of both the system models and the disturbance (e.g., load or renewable generations).}
Additionally, the computational cost of solving the OPF problem makes it difficult to respond to rapidly varying voltage profiles. 
\revise{On the other hand, feedback control methods do not assume to know the system model or the disturbance explicitly but take measurements of voltage magnitudes to decide the reactive power setpoints. In terms of time scale, the feedback controllers could work on a faster time scale as it does not require solving an optimization problem at each time step to decide the control actions.}
A popular feedback controller is the droop control, which is adopted by the IEEE 1547 standard \cite{8332112}. However, as shown in \cite{li2014real}, basic droop control can lead to instability if the controller gains are selected improperly. With more sophisticated structures, feedback controllers could achieve promising performance~\cite{qu2019optimal}. Regardless of the progress, optimizing the transient cost using feedback control methods remains challenging as the power flow equations are nonlinear, and the transient performance cost function can also be non-convex. The difficulty will be further exacerbated when the controllers are required to be optimized in a decentralized manner. 

\paragraph{Transient cost optimization} There has been tremendous interest in using RL for transient performance optimization in voltage control \cite{9356806,NEURIPS2021_1a672771,9353702,9328796,9143169,KOU2020114772,chen2021saver,8892476,wang_data-driven_2020,8944292,9113746,9274529,yeh2022robust}. Given different communication conditions, existing RL methods 
fall into three categories, centralized, distributed, and decentralized controllers. Centralized controllers mean that the agent has access to global operating conditions, which leads to a powerful controller~\cite{9328796}. However, sophisticated communication networks are demanded and the agent has to deal with high-dimensional information. In the distributed setting, the network is first partitioned into small regions, and each region is assigned with an RL agent~\cite{NEURIPS2021_1a672771,wang_data-driven_2020,9143169}. 
The agent has full observation of nodes located within the region.
Decentralized controllers\cite{9356806, shi2021stability,cui2020reinforcement} are trained only with local measurements, and thus require no communication among peers, reducing the local computational burden. Please refer to a recent review \cite{9721402} for more comprehensive overview. 
Despite the promise of RL for optimizing the transient performance, a widely-recognized issue is that RL lacks a provable stability guarantee, 
which is the main problem we want to tackle in this paper. 

\paragraph{Lyapunov approaches in RL} 
\revise{Using Lyapunov functions in RL was first introduced by \cite{perkins2002lyapunov}, where an agent learns to control the system by switching among the base controllers. These controllers are designed by using a specific Lyapunov function such that any switching policy is stable for the system. However, this work does not discuss how to find a candidate Lyapunov function in general, except for a case-by-case construction. A set of recent works including \cite{chow2018lyapunov,chang2019neural} have attempted to address this challenge by jointly learning a Lyapunov function and a stabilizing policy. \cite{chow2018lyapunov} uses linear programming to find the Lyapunov function, 
and \cite{chang2019neural} parameterizes the Lyapunov function as a neural network. To find a valid Lyapunov function and the corresponding controller, stability conditions are incorporated as a soft penalty during training and verified after training.}
In the context of these works, our contribution can be viewed as \emph{explicitly constructing} a Lyapunov function for the voltage control problem to guide policy learning, rather than  learning Lyapunov functions. Closest in spirit to our paper is \cite{cui2020reinforcement}, which
proposes a stable RL approach for frequency control. However, their approach only applies to
the frequency control problem, while our method works for
voltage control which requires a different Lyapunov function
design. Interestingly, both our work and prior work \cite{cui2020reinforcement} arrive
at a similar stability condition, that is strict policy monotonicity
guarantees system stability.

\section{Model \&  Preliminaries}
\label{sec:voltage}
In this section, we review distribution system power flow models for both single-phase and three-phase grids. 
\subsection{Branch Flow Model for Single-phase Grids}
\label{sec:voltage_control_formuation}
We consider the branch flow model~\cite{Baran1989a} in a radial distribution network. 
Consider a distribution grid $\mathcal{G} = (\mathcal{N}_0,\mathcal{E})$, consisting of a set of $\mathcal{N}_0 = \{0,1,\ldots,n\}$ nodes and an edge set $\mathcal{E}$.
In the graph, node $0$ is known as the substation, and all the other nodes are buses that correspond to residential areas. We also use $\mathcal{N} = \mathcal{N}_0/\{0\}$ to denote the set of nodes excluding the substation node. 
Each node $i\in\mathcal{N}$ is associated with an active power injection $p_i$ and a reactive power injection $q_i$. Let $V_i$ be the complex voltage and $v_i = |V_i|^2$ is the squared voltage magnitude. We use notation $\mathbf{p}, \mathbf{q}$ and $\mathbf{v}$ to denote the $p_i,q_i,v_i$ stacked into a vector. 
$\mathbf{p}, \mathbf{q}$ and $\mathbf{v}$ satisfy the following equations, $\forall j \in\mathcal{N}, i=\textrm{parent}(j)$,
\begin{subequations}
\label{eq:nonlinear_powerflow}
\begin{align}
    -p_j &= P_{ij} - r_{ij} {l_{ij}} - \sum_{k: (j, k) \in \mathcal{E}} P_{jk}, \label{eq:bfm_p}\\
    -q_j &= Q_{ij} - x_{ij} {l_{ij}} - \sum_{k: (j, k) \in \mathcal{E}} Q_{jk}, \label{eq:bfm_q}\\
    v_j &= v_i - 2(r_{ij}P_{ij} + x_{ij} Q_{ij}) + (r_{ij}^2 + x_{ij}^2) l_{ij}, (i,j)\in \mathcal{E}  \label{eq:bfm_v} 
\end{align}
\end{subequations}
where $l_{ij} = \frac{P_{ij}^2 + Q_{ij}^2}{v_i}$ is the squared current, $P_{ij}$ and $Q_{ij}$ represent the active power and reactive power flow on line $(i,j)$, and $r_{ij}$ and $x_{ij}$ are the line resistance and reactance. 
Equation~\eqref{eq:bfm_p} and \eqref{eq:bfm_q} represent the real and reactive power conservation at node $j$, and \eqref{eq:bfm_v} represents the voltage drop from node $i$ to node $j$. 

Following \cite{baran1989network}, if the higher order power loss term can be ignored by setting $l_{ij} = 0$, we obtain the following linear approximation model,
 \begin{subequations}\label{eq:linear_distflow}
{\small
\begin{align}
    p_j &= -P_{ij}  + \sum_{k: (j, k) \in \mathcal{E}} P_{jk}\,, \quad
    q_j = -Q_{ij} + \sum_{k: (j, k) \in \mathcal{E}} Q_{jk}\,, \label{eq:conserv_law}\\
    v_j &= v_i - 2(r_{ij}P_{ij} + x_{ij} Q_{ij})\,, (i, j) \in \mathcal{E} \label{eq:bfm_vl}
\end{align}}
\end{subequations}
We can rearrange the above equations into the vector form, 
\begin{equation}
\mathbf{v} = R \mathbf{p} + X \mathbf{q} + v_0 \mathbf{1} = X\mathbf{q} + \mathbf{v}^{env}. \label{eq:lindistflow1}
\end{equation}
where matrix $R={[R_{ij}]}_{n \times n}, X = {[X_{ij}]}_{n \times n}$ are given as follows,
$R_{ij}:= 2 \sum_{(h, k) \in \mathcal{P}_i \cap \mathcal{P}_j} r_{hk}, X_{ij}:= 2 \sum_{(h, k) \in \mathcal{P}_i \cap \mathcal{P}_j} x_{hk}$ where $\mathcal{P}_i \subset E$ is the set of lines on the unique path from bus $0$ to bus $i$. 
Here we follow~\cite{li2014real} to separate the voltage magnitude $\mathbf{v}$ into two parts: the controllable part $X\mathbf{q}$ that can be adjusted via adjusting reactive power injection $\mathbf{q}$ through the inverter-based control devices, and the non-controllable part $\mathbf{v}^{env} = R\mathbf{p}+v_0 \mathbf{1}$ that is decided by the load and PV active power $\mathbf{p}$.  
Matrix $X$ and $R$ satisfy the following property, which is crucial for the stable control design. 
\begin{proposition}[\cite{li2014real} Lemma 1]\label{prop:RX_pd}
Suppose $x_{ij}, r_{ij}>0$ for all $(i,j)$. Then, $X$ and $R$ are positive definite matrices. 
\end{proposition}

\subsection{Multi-phase Grid Modeling}
We now introduce an abridged version of the branch flow model in three-phase distribution systems. For simplicity, it is first assumed that all buses are served by all three phases, so we can use 3-dimensional vectors to represent system variables. With slight abuse of notation, $\mathbf{P_{ij}}$ is a 3-dimensional vector such that $\mathbf{P_{ij}}=[P_{ij}^a,P_{ij}^b,P_{ij}^c]^\top$. $\mathbf{S_{ij}}$, $\mathbf{Q_{ij}}$ are defined in the same way. The vectors of power injections and complex voltages are denoted by $\mathbf{s_i}$ and $\mathbf{v_i}$, respectively. 
$\mathbf{Z_{ij}}\in \mathbb{S}^3$ is the phase impedance matrix for line $(i,j)$, where $\mathbf{Z_{ij}}=\mathbf{R_{ij}}+j\mathbf{X_{ij}}$. 

We further assume that the phase voltages of arbitrary bus $i$ are approximately balanced with absolute value $\hat{v}_i$, then $\mathbf{v_i}$ can be estimated by $\hat{v}_i\mathbf{\alpha}$, where $\mathbf{\alpha} = [1\quad a\quad a^2]$, $a = e^{-j{2\pi}/{3}}$. Define $\mathbf{\hat{Z}_{ij}}=diag(\mathbf{\alpha}^H)\mathbf{Z_{ij}}diag(\mathbf{\alpha})$, following \eqref{eq:linear_distflow}, the linear approximate three-phase model is, 
\begin{subequations}
\begin{align}
    \mathbf{s_j}&=-\mathbf{S_{ij}} + \sum_{k: (j, k) \in \mathcal{E}} \mathbf{S_{jk}}\\
    \mathbf{v_i}-\mathbf{v_j}&=2Re[\mathbf{\hat{Z}_{ij}}^H\mathbf{S_{ij}}]
\end{align}
\end{subequations}
Notice that the vector variables can be arranged either by bus or by phase. For example, the voltage magnitude can be rearranged by phase as $\mathbf{\check{v}}=[\mathbf{\check{v}_a},\mathbf{\check{v}_b},\mathbf{\check{v}_c}]$, where $\mathbf{\check{v}_a} = [v_1^a,v_2^a,...,v_n^a]$, $\mathbf{\check{v}_b}$ and $\mathbf{\check{v}_c}$ share the similar definition. Recall that $\mathbf{v}=[\mathbf{v_i}, i \in\mathcal{N}]$, which is ordered by bus. With a permutation matrix $T_v$, the transformation between two formats can be represented by $\mathbf{v}= T_v \mathbf{\check{v}}$. The three-phase branch flow model can then be arranged to a compact form, the same as the single-phase model,
\begin{equation}
\mathbf{v} = R \mathbf{p} + X \mathbf{q} + v_0 1_{3N} = X\mathbf{q} + \mathbf{v}^{env}. \label{eq:lindistflow}
\end{equation}
For a detailed mathematical derivation, please refer to \cite{7317598}. Notice that the single-phase and three-phase system dynamics share the same linear approximation model~\eqref{eq:lindistflow1}, $\mathbf{v} = X\mathbf{q} + \mathbf{v}^{env}$, allowing us to derive the Lyapunov equation and stability conditions based on the same analytical model.

\revise{
\assumption{Assume every matrix $\mathbf{X_{ij}} = \frac{1}{2} \begin{bmatrix} 2x^{aa}_{ij} & -x^{ab}_{ij} & x^{ac}_{ij}\\
-x^{ab}_{ij} & 2x^{bb}_{ij} & x^{bc}_{ij}\\
-x^{ac}_{ij} & -x^{bc}_{ij} & 2x^{cc}_{ij}\end{bmatrix}$ is strictly diagonally dominant with positive diagonal entries for all edges $(i, j) \in \mathcal{E}$.}\label{assmp:1}
}

\revise{It is denoted in~\cite{7317598}, due to the structure of distribution lines, the diagonal dominance conditions in Assumption \ref{assmp:1} are generally satisfied for multi-phase grids. According to Corollary 1 in~\cite{7317598}, if Assumption \ref{assmp:1} holds, $X$ is positive definite for three-phase distribution system.}

\section{Voltage Control Problem Formulation}
The voltage control problem can be modeled as a control problem in a quasi-dynamical system with state $\mathbf{v}$ and controller $\mathbf{q}$. Given the current voltage measurement $\mathbf{v}(t)$ and other available information, the controller determines a new reactive power injection $\mathbf{q}(t+1)$. The new $\mathbf{q}(t+1)$ will result in a new voltage profile $\mathbf{v}(t+1)$. We envision that the reactive power loop is embedded in an inverter control loop and operates at very fast timescales~\cite{9328796}, and denote the change rate of reactive power injection as, $\dot{q}_i(t):=u_i(t)$. Using the zero-order hold on the inputs and a sample time of $\Delta T$, we get the closed-loop voltage control dynamics as,
\begin{subequations}
\begin{align}
   \mathbf{v}(t+1) &= X\mathbf{q}(t+1) + \mathbf{v}^{env} \,,\\
   q_i(t+1) &= q_i(t) + \Delta T \cdot u_i(v_i (t)), \forall i \in \mathcal{N} \label{eq_controller_discrete}
\end{align}
\end{subequations}
where $\mathbf{u} = (u_1, \cdots, u_n)$ is the decentralized voltage controller. 
Note that \eqref{eq_controller_discrete} represents the class of \emph{incremental voltage controller}. 
As shown in \cite{li2014real}, a decentralized controller only depends on the current time step information, i.e., $q_i(t) = u_i(v_i(t))$ is \emph{not possible} to stabilize the voltage $\bf{v}$ under arbitrary disturbance, while the incremental voltage controller guarantees the existence of stabilizing controllers. This motivates our focus on incremental voltage controllers.

\subsection{Voltage Stability}
Voltage stability is defined as the ability of the system voltage trajectory to return to an acceptable range \emph{after arbitrary disturbance}. See Definition~\ref{def:voltage_stability} below.
\revise{\begin{definition}[Voltage stability]
\label{def:voltage_stability}
The closed loop system is stable if for any $\mathbf{v}^{env}$ and $\mathbf{v}(0)$,  $\mathbf{v}(t)$ converges to the set $S_{v} = \{ \mathbf{v}\in\mathbb{R}^n: \underline{v}_i\leq v_i \leq\bar{v}_i \}$ in the sense that $\lim_{t\rightarrow\infty }\dist(\mathbf{v}(t),S_v) = 0$ and the distance is defined as $\dist(\mathbf{v}(t),S_v) = \min_{\mathbf{v}' \in S_v}||\mathbf{v}(t)-\mathbf{v}'||$.
\end{definition}
\begin{figure}[h]
    \centering
    \includegraphics[width=0.48\textwidth]{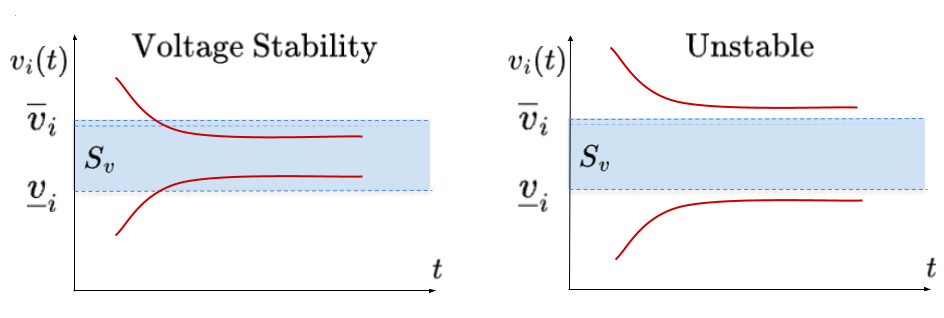}
    \caption{\revise{Voltage stability of bus i.}}
    \label{fig:voltage stability}
\end{figure}}
\begin{figure}[t]
    \centering
    \includegraphics[width=0.45\textwidth]{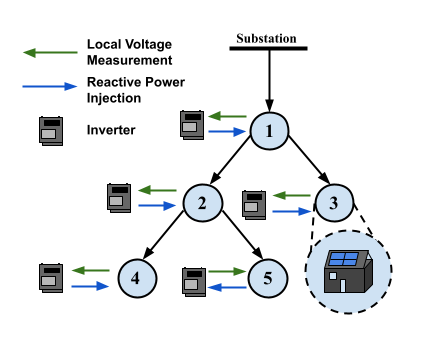}
    \caption{Control System Architecture.}
    \label{fig:decentralised controller}
\end{figure}
With high penetration of DERs, rapid changes of load and renewable generation often happen in a fast time scale, thus it is important to ensure the designed controller meets the stability condition. With the requirement for voltage stability, the optimal voltage control problem can be formulated as, 
\begin{subequations}
\label{opt:voltage_ctrl-discrete}
\begin{align}
\min_{\mathbf{\theta}} \quad & J(\theta)= \sum_{t=0}^{T} \gamma^t \sum_{i=1}^n c_i({v}_i(t), u_i(t)) \label{eq_rl:obj-dis}\\
\text{s.t.}\quad 
& \mathbf{v}(t+1) = X \mathbf{q}(t+1) + \mathbf{v}^{env}, \label{eq_rl:dyn-dis1}\\ 
& q_i(t+1)=q_i(t) + \Delta T \cdot u_i(t) \label{eq_rl:dyn-dis2}\\
&u_i(t) = -g_{\theta_i}(v_i(t))\label{eq_rl:policy-dis}\\
&\text{Voltage stability holds.}\label{eq_rl:constr}
\end{align}
\end{subequations}
The goal of the voltage control problem is to reduce the total cost \eqref{eq_rl:obj-dis} for time steps $t$ from 0 to $T$, which consists of two parts: the cost of voltage deviation and the cost of control actions. One can choose different cost functions (e.g., one-norm, two-norm, or infinity-norm), depending on the system performance metrics and control devices. Our stability-constrained RL framework can accommodate different types of cost functions mentioned above. 
In particular, in our experiment, we use $c_i({v}_i(t), u_i(t)) = \eta_1 \|\max(v_i(t) - \bar{v}_i,0) + \min(v_i(t) - \underline{v}_i,0)\|_{2}^2 + \eta_2 \|u_i(t)\|_1$.
Here $\eta_1, \eta_2$ are coefficients that balance the cost of action with respect to the voltage deviation.  
Voltage dynamics are represented by equations~\eqref{eq_rl:dyn-dis1}-\eqref{eq_rl:dyn-dis2}. 
\eqref{eq_rl:policy-dis} specifies the decentralized policy structure $u_i(t) = -g_{\theta_i}(v_i(t))$ only depends on local voltage measurement $v_i(t)$. 
Here $\theta_i$ is the policy parameter for the local policy at node $i$, and $\theta = (\theta_i)_{i\in\mathcal{N}}$ is the collection of the local policy parameters.


\textit{Transient cost vs. stationary cost. } Our problem formulation in \eqref{opt:voltage_ctrl-discrete} is different from some of those in the literature, e.g., \cite{bolognani2013distributed,zhu2016fast,tang2019fast,qu2019optimal}, in the sense that the existing works typically consider the cost in steady-state, meaning the cost is evaluated at the fixed point or stationary point of the system. In contrast, our work considers the transient cost after a voltage disturbance, which is also an important metric for the performance of voltage control.
An important future direction is to unify these two perspectives and design policies that can optimize both the transient and stationary costs. 

\subsection{Solving Voltage Control Problem via RL}
\label{sec:ddpg_voltagectrl}
In order to solve the optimal voltage control problem in \eqref{opt:voltage_ctrl-discrete}, one needs the exact system dynamics, i.e., $X$. However, for distribution systems, the exact network parameters are often unknown or hard to estimate in real systems~\cite{chen_input_2020}. 
RL provides a powerful paradigm for solving \eqref{opt:voltage_ctrl-discrete}, by training a policy that maps the state to action via interacting with the environment, so as to minimize the loss function defined as~\eqref{eq_rl:obj-dis}. 
There are many RL algorithms to solve the policy minimization problem \eqref{opt:voltage_ctrl-discrete}, and in this paper, we focus on the class of RL algorithms called \emph{policy optimization}. We define the state space of each local controller as the nodal voltage deviation, represented by $v_i \in \mathbb{R}$ (single-phase) or $v_i \in \mathbb{R}^3$ (three-phase). The action space is defined as the range of potential reactive power changes, represented by $ u_i\in\mathbb{R}$ (single-phase) or $u_i \in \mathbb{R}^3$ (three-phase).
%

Generally speaking, we parameterize each of the controllers, i.e., $u_i(t) = -g_{\theta_i}(v_i(t))$ as a neural network with weights $\theta_i$. The procedure is to run gradient methods on the policy parameter $\theta_i$ with learning rate $\alpha_i$,
$ \theta_i \leftarrow \theta_i - \alpha_i \nabla J(\theta_i). $
As we are dealing with deterministic policies and continuous state space, one of the most popular choices is the Deep Deterministic Policy Gradient (DDPG)~\cite{lillicrap2015continuous}, where the policy gradient $\nabla J(\theta_i)$ is approximated by 
\begin{equation}
\label{eq:policy_update}
-\frac{1}{N} \sum_{j\in B} \nabla_{u_i} \hat{Q}_{\phi_i}(v_i, u_i)|_{v_i = v_i[j], u_i=-g_{\theta_i}(v_i[j])} \nabla_{\theta_i} g_{\theta_i}(v_i)|_{v[j]}\,,
\end{equation}
where $g_{\theta_i}(v_i)$ is the actor network, and $\{v_i[j], u_i[j]\}_{j\in B}$ are a batch of samples with batch size $|B| = N$ sampled from the replay buffer which stores historical state-action transition tuples of bus $i$. 
Here $\hat{Q}_{\phi_i}(v_i, u_i)$ is the value network (a.k.a critic network) that can be learned via temporal difference learning,
\begin{equation}
    \label{eq:value_update}
    \min_{\phi_i} L(\phi_i) = E_{(v_i, u_i, c_i, v_i')} [Q_{\phi_i}(v_i, u_i) - (c_i+\gamma Q_{\phi_i}(v_i', g_{\theta_i}(v_i'))]
\end{equation}
where $v_i'$ is system voltage after taking action $u_i$ and realization of $v^{env}_i$.
For more details of DDPG, readers may refer to~\cite{lillicrap2015continuous}.

In standard DDPG, stability is not an explicit requirement, it plays the role of implicit regularization since instability leads to high costs. However, the lack of an explicit stability requirement can lead to several issues. During the training phase, the policy may become unstable, causing the training process to terminate. Even after a policy is trained, there is no formal guarantee that the closed loop system is stable, which hinders the learned policy's deployment in real-world power systems where there is a very strong emphasis on stability. Next, we will introduce our framework that guarantees stability in policy learning.

\vspace{-6pt}
\section{Main Results}
\label{sec:method}
We now introduce our stability-constrained RL framework for voltage control. 
We demonstrate that the voltage stability constraint can be translated into a monotonicity constraint on the policy, that can be satisfied by a careful design of monotone neural networks.
\vspace{-6pt}
\subsection{Voltage Stability Condition} 
\label{sec:mono}
In order to explicitly constrain stability for RL, we constrain the search space of policy in a subset of stabilizing controllers from Lyapunov stability theory. In particular, we use a generalization of Lyapunov's direct method, known as LaSalle's Invariance theorem for deriving the stability condition.
\begin{proposition}[LaSalle's theorem for discrete-time system~\cite{bof2018lyapunov}]
\label{prop:discrete_sys_lasalle}
For dynamical system $x(t+1) = f(x(t))$, suppose $V:\mathbb{R}^n\rightarrow \mathbb{R}$ is a continuously differentiable function satisfying $V(x)\geq 0$ and $V(f(x)) - V(x) \leq 0, \forall x\in\mathbb{R}^n$. Let $E$ be the set of all points in $\mathbb{R}^n$ where $V(f(x)) - V(x) = 0$, and let $M$ be the largest invariant set in $E$. If there exists $a\in\mathbb{R}^{+}$ such that the level set $L_a:=\{x: V(x)\leq a\}$ is bounded, then for any $x(0)\in L_a$ we have $\dist(x(t),M)\rightarrow 0$ as $t \rightarrow \infty$. Further, if $V$ is radially unbounded, i.e. $V(x)\rightarrow\infty$ as $\Vert x\Vert\rightarrow\infty$, then, for any $x(0)\in\mathbb{R}^n$, we have $\dist(x(t),M)\rightarrow 0$ as $t \rightarrow \infty$.
\end{proposition}
%


The key to ensure stability is to find a controller $\mathbf{u} =-g_{\theta}(\mathbf{v})$ and a Lyapunov function $V$, such that the stability conditions in Proposition~\ref{prop:discrete_sys_lasalle} can be satisfied.
For the voltage control problem defined by \eqref{opt:voltage_ctrl-discrete}, $\mathbf{v}(t+1) = \mathbf{v}(t) + I_{\Delta T}X\mathbf{u}(t)$, where $I_{\Delta T}$ is a diagonal matrix with diagonal entries equal to $\Delta T$. Since the control input $\mathbf{u} = -g_{\theta}(\mathbf{v})$ depends on state $\mathbf{v}$, the closed-loop system dynamics can be written as $\mathbf{v}(t+1)  = \mathbf{v}(t) - I_{\Delta T} X g_{\theta}(\mathbf{v}) := f_{u}(\mathbf{v}(t))$.
We consider the following Lyapunov function, 
\begin{equation}\label{eq:vol_lya}
    V(\mathbf{v}) = (\mathbf{v}-f_{u}(\mathbf{v}))^\top X^{-1} (\mathbf{v}-f_{u}(\mathbf{v}))
\end{equation}
where X is the network reactance matrix defined in \eqref{eq:lindistflow1}, that is a positive definite matrix for both single-phase and three-phase distribution grids. $V$ is positive definite and is radially unbounded if $\|g_{\theta}(\mathbf{v})\| \rightarrow \infty$ as $\|\mathbf{v}\| \rightarrow \infty$.
From LaSalle's theorem in Proposition~\ref{prop:discrete_sys_lasalle}, if $V(f_{u}(\mathbf{v}))-V(\mathbf{v})\leq 0$ and  $V(f_{u}(\mathbf{v}))-V(\mathbf{v}) = 0$ only when $\mathbf{v}\in S_v$, where $S_v$ is the voltage safety set defined as $S_{v} = \{ \mathbf{v}\in\mathbb{R}^n: \underline{v}_i\leq v_i \leq\bar{v}_i \}$. Then we have for every initial voltage profile $\mathbf{v}(0) \in \mathbb{R}^n$, $\mathbf{v}(t)$ converges to the largest invariant set in $S_{v}$. 
Furthermore, suppose for all $i$, the control action satisfies $u_i = 0$ for $v_i \in [\underline{v}_i,\bar{v}_i]$. Then $S_v$ itself is an invariant set.

The key question now reduces to how can we design the controller $\mathbf{u} = -g_{\theta}(\mathbf{v})$ such that the closed-loop system satisfying these two properties: 
\begin{enumerate}
    \item $V(f_{u}(\mathbf{v}))-V(\mathbf{v}) < 0$, $\forall v \notin S_v$
    \item $V(f_{u}(\mathbf{v}))-V(\mathbf{v}) = 0$ for $\mathbf{v}\in S_v$
\end{enumerate}
Theorem \ref{thm:voltage_stab} presents a sufficient structural condition for the above properties to hold, thus guaranteeing voltage stability.
\begin{theorem}[Voltage stability condition]\label{thm:voltage_stab} 
Suppose for all bus $i$, $g_{\theta_i}(\cdot)$ is a continuously differentiable function satisfying $u_i = -g_{\theta_i}(v_i) = 0$ for $v_i \in [\underline{v}_i,\bar{v}_i]$. Further, each $\frac{\partial g_{\theta_i}}{\partial v_i}$ satisfies equation \eqref{eq:stability_cond_discrete} on $(-\infty,\underline{v}_i]$ and $[\bar{v}_i,\infty)$
\begin{equation}\label{eq:stability_cond_discrete}
   -\frac{2}{\Delta T} X^{-1}  \prec \frac{\partial \mathbf{u}}{\partial \mathbf{v}} \prec 0
\end{equation}
and $\lim_{|v_i|\rightarrow\infty} |g_{\theta_i}(v_i)| = \infty$.
Then, the voltage stability defined in Definition~\ref{def:voltage_stability} holds.
\end{theorem}
Equation \eqref{eq:stability_cond_discrete} shows that when the sampling time $\Delta T\to 0$, the stability condition reduces to the continuous time stability condition $\frac{\partial \mathbf{u}}{\partial \mathbf{v}} \prec 0$ as first shown in \cite{shi2021stability}. As the length of the sampling time increases, $\frac{\partial \mathbf{u}}{\partial \mathbf{v}}$ also needs to be lower bounded by $- \frac{2}{\Delta T} X^{-1}$ and upper bounded by $0$. As the typical sampling frequency of real-world inverters is in kHz scale\cite{en14165170}, the left-hand side of \eqref{eq:stability_cond_discrete} is naturally satisfied in most cases. 
Therefore, we focus on the right-hand side condition, $\frac{\partial \mathbf{u}}{\partial \mathbf{v}} \prec 0$. Because of the decentralized characteristic, $\frac{\partial \mathbf{u}}{\partial \mathbf{v}}$ is a diagonal matrix. \begin{equation}
    \frac{\partial \mathbf{u}}{\partial \mathbf{v}} = -\begin{bmatrix}\frac{\partial  g_{\theta_1}}{\partial v_1}  & \cdots & 0\\
     & \ddots  & \\
    0 &  \cdots & \frac{\partial  g_{\theta_n}}{\partial v_n}
    \end{bmatrix}
\end{equation}
Thus, if each $g_{\theta_i}$ is strictly monotonically increasing, i.e., $\frac{\partial g_{\theta_i}(v_i)}{\partial v_i} > 0, \forall i$, the voltage stability condition $\frac{\partial \mathbf{u}}{\partial \mathbf{v}} \prec 0$ will be met. 
We note that a similar stability condition for the discrete-time voltage dynamics has been shown in~\cite{cui2021decentralized}, while our condition ensures globally asymptotic stability 
rather than the local stability guarantee in~\cite{cui2021decentralized}.

\subsection{Stability-Constrained RL Algorithm}
\begin{algorithm}[t]
	\caption{Stable-DDPG Learning Process}
	\label{alg:stable_ddpg}
	\begin{algorithmic}[1]
	\Ensure Initial Q network $Q_{\phi_i}(v_i,u_i)$, \emph{monotone policy network} $g_{\theta_i}(v_i)$ with parameters $\phi_i, \theta_i$; empty replay buffers $\mathcal{D}_i$ for all buses $i = 1,..., n$. 
	\For {$j = 0$ to $N_{ep}$}
	  \State{Randomly generate initial states $\mathbf{v}(0)$ with voltage violation for all nodes}
	  \For {$t = 0$ to $N_{step}$}
	    \State{Observe state $v_i(t)$, compute action based on current policy $u_i = -g_{\theta_i}(v_i(t))$, $\forall i$}
	    \State {Execute the joint action $\mathbf{q}(t+1) = \mathbf{q}(t) + \Delta T \mathbf{u}(t)$, transit to next state $\mathbf{v}(t+1)$}
	    \State{Store $(v_i(t),u_i(t),c_i(t),v_i(t+1))$ in $\mathcal{D}_i$, $\forall i$ }
	  \EndFor
	  \If{len($\mathcal{D}_i$) $> $ batch size}
	  \For {$i=1,...,n$}
	    \State{Randomly sample $N$ state-transition data pairs from replay buffer $D_i$, $B_i = \{(v_i,u_i,c_i,v_i')\}_{i=1}^{N}$.}
	    \State{Update the policy network by Eq~\eqref{eq:policy_update}}
	    \State{Update the Q-function network Eq~\eqref{eq:value_update}}
	   \EndFor
	  \EndIf
	\EndFor
	\end{algorithmic}
\end{algorithm}
Combining the structural constraints for stabilizing controllers in Theorem \ref{thm:voltage_stab} and the DDPG algorithm for solving voltage control in Section \ref{sec:ddpg_voltagectrl}, we now present the design of Stable-DDPG algorithm. The proposed stability-constrained policy learning algorithm is summarized in Algorithm~\ref{alg:stable_ddpg}.

As we notice in Algorithm~\ref{alg:stable_ddpg}, the general algorithm flow of Stable-DDPG is the same as DDPG, and the only difference is in the policy network parameterization. Since Theorem \ref{thm:voltage_stab} restricts the class of stabilizing decentralized controllers to be \emph{strictly monotonically decreasing}. Thus, we need to incorporate this structural condition into the policy design. Essentially, any monotone functions can be used for parameterizing the policy function, e.g., a linear policy $u_i = -k_i v_i, \forall v_i < \underline{v}_i$ or $v_i > \overline{v}_i$ with $k_i$ is positive; and $u_i = 0$ for $\underline{v}_i \leq v_i \leq \overline{v}_i$. To leverage the superior expressiveness of neural networks, we represent $u_i = -g_{\theta_i}(v_i)$ as a monotone neural network. 
There are several existing designs for the monotone neural networks in literature, e.g.  \cite{wehenkel2019unconstrained,cui2023structured,cui2020reinforcement}.  
In this paper, we follow the monotonic neural network design in~\cite[Lemma 3]{cui2020reinforcement}, \revise{which guarantees universal approximation of all single-input-single-output monotonic increasing functions \cite[Theorem 2]{cui2020reinforcement}. 
This design uses a single hidden layer neural network with $d$ hidden units and ReLU activation, which is defined below.}
\begin{corollary}(Stacked ReLU Monotone Network~\cite[Lemma 3]{cui2020reinforcement})
\label{corollary:stacked_relu}
 The stacked ReLU function constructed by Eq~\eqref{eq:relu_pos} is monotonic increasing for $x > 0$ and zero when $x \leq 0$.
\begin{small}
\begin{subequations}\label{eq:relu_pos}
    \begin{align}
    &\xi^{+}(x; w^+, b^+) = {(w^+)^\top} \text{ReLU}(\mathbf{1} x + b^+)\\
    &\sum_{l=1}^{d'} w^{+}_l > 0, \forall d' = 1, ..., d\,, b^{+}_1 = 0, b^{+}_l \leq b^{+}_{l-1}, \forall l =2, ..., d
    \end{align}
\end{subequations}
\end{small}
The stacked ReLU function constructed by Eq~\eqref{eq:relu_neg} is monotonic increasing for $x < 0$ and zero when $x \geq 0$.
\begin{small}
\begin{subequations}\label{eq:relu_neg}
    \begin{align}
    &\xi^{-}(x; w^{-}, b^{-}) = (w^{-})^\top \text{ReLU}(-\mathbf{1} x + b^{-})\\
    & \sum_{l=1}^{d'} w^{-}_l < 0, \forall d' = 1, ..., d\,, b^{-}_1 = 0, b^{-}_l \leq b^{-}_{l-1}, \forall l =2, ..., d
    \end{align}
\end{subequations}
\end{small}
\end{corollary}

\subsubsection{Single-phase Monotone Voltage Controller}
\label{sec:single-phase}
Following the stability constraint \eqref{eq:stability_cond_discrete}, we set the single-phase voltage controller to be monotonically increasing with Corollary~\ref{corollary:stacked_relu}.
To incorporate the dead-band within range $v_i \in [\underline{v}_i, \overline{v}_i]$, we parameterize the controller at bus $i$ as $g_{\theta_i}(v_i) = [\xi_{\theta_i}^{+}(v_i-\overline{v}_i) + \xi_{\theta_i}^{-}(v_i-\underline{v}_i)]$, where $\xi_{\theta_i}^{+}(v_i-\overline{v}_i): \mathbb{R} \rightarrow \mathbb{R}$ is monotonically increasing for $v_i >\overline{v}_i$ and zero when $v_i \leq \overline{v}_i$, and $\xi_{\theta_i}^{-}(v_i-\underline{v}_i): \mathbb{R} \rightarrow \mathbb{R}$ is monotonically increasing for $v_i <\underline{v}_i$ and zero otherwise. Because $u_i(t) = -g_{\theta_i}(v_i(t))$, $\frac{\partial \mathbf{u}}{\partial \mathbf{v}} \prec 0$ is satisfied. 

\subsubsection{Three-phase Monotone Voltage Controller}
\label{sec:three-phase}
For the three-phase voltage controller, we generalize the single-in-single-out monotone policy network to three-dimensional input and output by deploying a single-phase controller for each phase. 
\begin{figure}[t]
    \centering
    \includegraphics[width=0.45\textwidth]{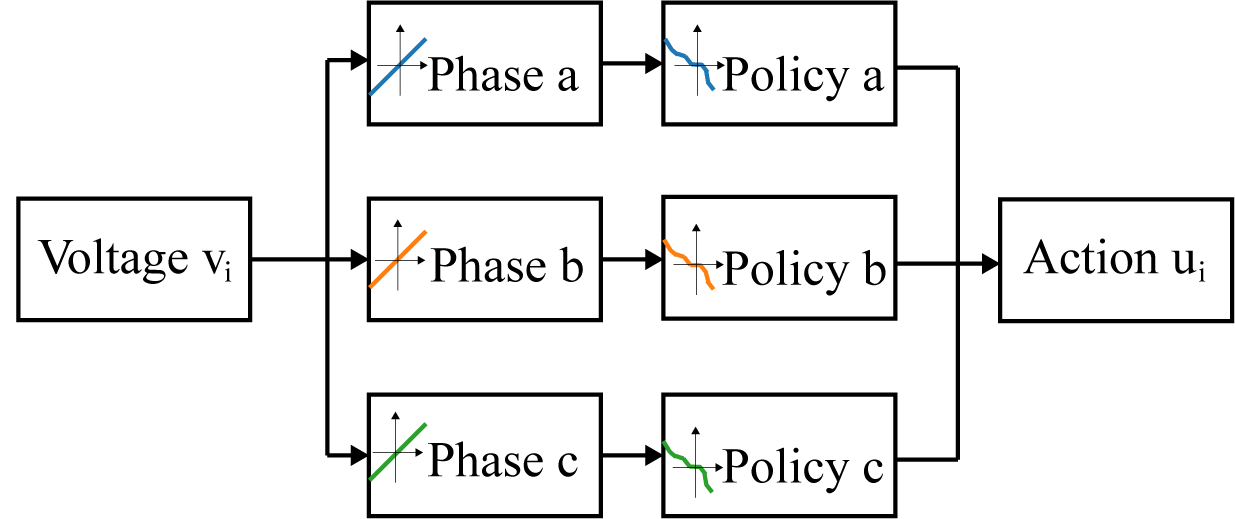}
    \caption{Three-phase Monotone Voltage Controller, where each policy is parameterized as a Stacked ReLU Monotone Network.}
    \label{fig:three-phase-controller}
\end{figure}
As demonstrated in Figure \ref{fig:three-phase-controller}, we disentangle the AC voltage observations per phase and treat each phase as a single-phase input. In this way, $\frac{\partial u_i}{\partial v_i}=\left [ \begin{matrix}\frac{\partial u_i^a}{\partial v_i^a}&0&0\\0&\frac{\partial u_i^b}{\partial v_i^b}&0\\0&0&\frac{\partial u_i^c}{\partial v_i^c}\end{matrix}\right ]$ is a diagonal matrix with negative entries, thus the stability condition $\frac{\partial \mathbf{u}}{\partial \mathbf{v}} \prec 0$ is satisfied. 

We conclude this section with two remarks. 
\begin{remark}
The simplified Distflow models in \eqref{eq:lindistflow1} and \eqref{eq:lindistflow} are introduced for theoretic analysis only. The original nonlinear dynamics are deployed in numerical experiments, and our proposed method stabilizes the systems in all test scenarios. 
\end{remark}

\begin{remark}
The stability criteria defined by Theorem \ref{thm:voltage_stab} is a sufficient condition for asymptotic stability, which does not give any explicit guarantee to stabilize the systems in finite steps. To achieve exponential stability, the Lyapunov condition should be strengthened as $V(\mathbf{v}_{t+1})-V(\mathbf{v}_t)\leq -cV(\mathbf{v}_t),$ $0<c<1$ \cite{bof2018lyapunov}. In this case, the stability condition is given by $-\frac{1+\sqrt{1-c}}{\Delta T} X^{-1}  < \frac{\partial \mathbf{u}}{\partial \mathbf{v}} < -\frac{1-\sqrt{1-c}}{\Delta T}X^{-1}$. 
{Given that $\Delta T$ is small, we can often find a constant $c$ such that the trained policy satisfies the above inequality. As a result, the system can be input-to-state stable~\cite{khalil2002nonlinear} with the proposed controller. }
\end{remark}

\section{Case Study}
\label{sec:experiment}
We demonstrate the effectiveness of the proposed Stable-DDPG approach (Algorithm~\ref{alg:stable_ddpg}) on both single-phase and three-phase IEEE distribution test systems. Source code and data are available at \href{https://github.com/JieFeng-cse/Stable-DDPG-for-voltage-control}{https://github.com/JieFeng-cse/Stable-DDPG-for-voltage-control}.  
\subsection{Experimental Setup}
For single-phase test feeders, we use the IEEE 13-bus feeder and IEEE 123-bus feeder as the test cases, which are modified from three-phase models in \cite{8063903}. 
For three-phase test feeders, we test on the unbalanced IEEE 13-bus feeder \revise{and IEEE 123-bus feeder}~\cite{8063903}. Simulations for single-phase systems are implemented with pandapower~\cite{pandapower}, and the three-phase system is obtained by OpenDSS. We simulate different voltage disturbance scenarios: 1) High voltages: the PV generators are generating a large amount of power, this corresponds to the daytime scenario in California where there is abundant sunshine that can result in high voltage issues. 2) Low voltages: the system is serving heavy loads without PV generation. It corresponds to late afternoon or night when there is low/no solar generation but still a significant load. For each scenario, we randomly vary the active power injections 
thus obtaining different degrees of voltage violations, i.e., $5\%$ to $15\%$ of the nominal value. \revise{We set $\Delta T=1s$ for all numerical experiments and verify the stability condition for all the trained policies. All experiments are conducted with an AMD 5800X CPU and an Nvidia 1080Ti GPU.}

\textbf{Baselines:} We test the proposed stable-DDPG approach (Algorithm~\ref{alg:stable_ddpg}), against the following baseline algorithms. Details about the algorithm implementations are provided in Appendix B.

\paragraph{Linear policy with deadband} $u_i(v_i) = -\epsilon_i ([v_i - \overline{v}_i]^{+} - [\underline{v}_i - v_i]^{+})$ (where $[x]^+= \max(x,0)$), and the new reactive power injection is $q_i(t) = q_i(t-1) + \Delta T u_i(v_i)$. This linear controller has been widely used in the power system control community~\cite{li2014real}. With each $0< \epsilon_i \leq \frac{2\sigma_{min}(X)}{\sigma_{max}^2(X)}$, linear policy guarantees stability but may lead to suboptimal control cost. We optimize the linear controller in an RL framework, where $\epsilon_i$ is the learnable parameter to obtain the best-performing linear policy for comparison. 

\paragraph{Standard DDPG algorithm} is suggested for voltage control in~\cite{wang_data-driven_2020,KOU2020114772}. Standard DDPG minimizes the control cost without an explicit stability guarantee. 

\revise{\paragraph{DDPG*} We denote the \emph{subset} of results where the standard DDPG policy is able to maintain voltage stability as DDPG*. }

\subsection{Single-phase Simulation Results}
\subsubsection{13-bus system}
\begin{figure}[t]
    \centering
    \includegraphics[width=0.24\textwidth]{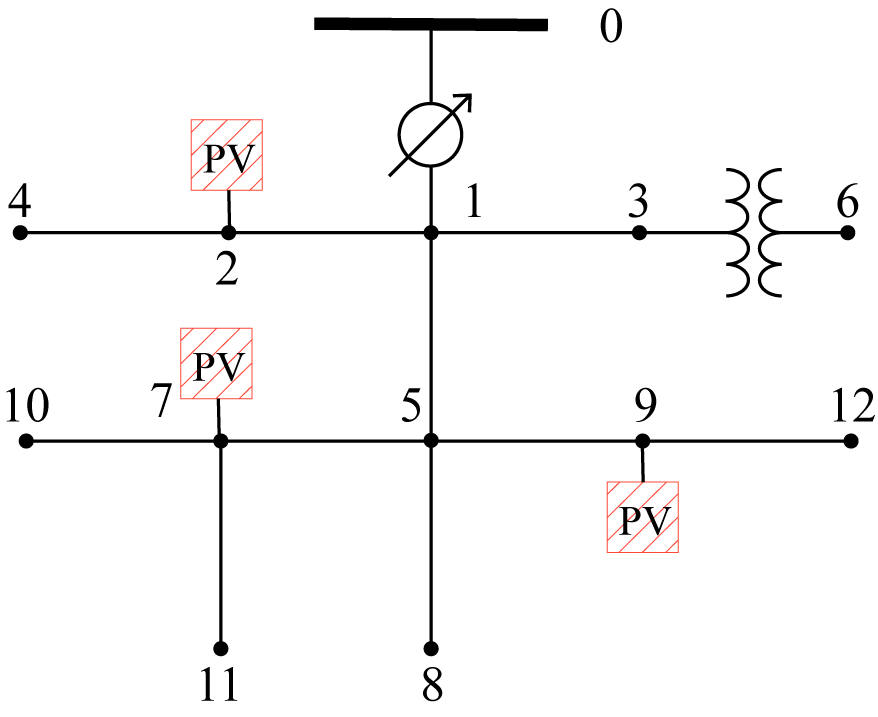}
    \includegraphics[width=0.24\textwidth]{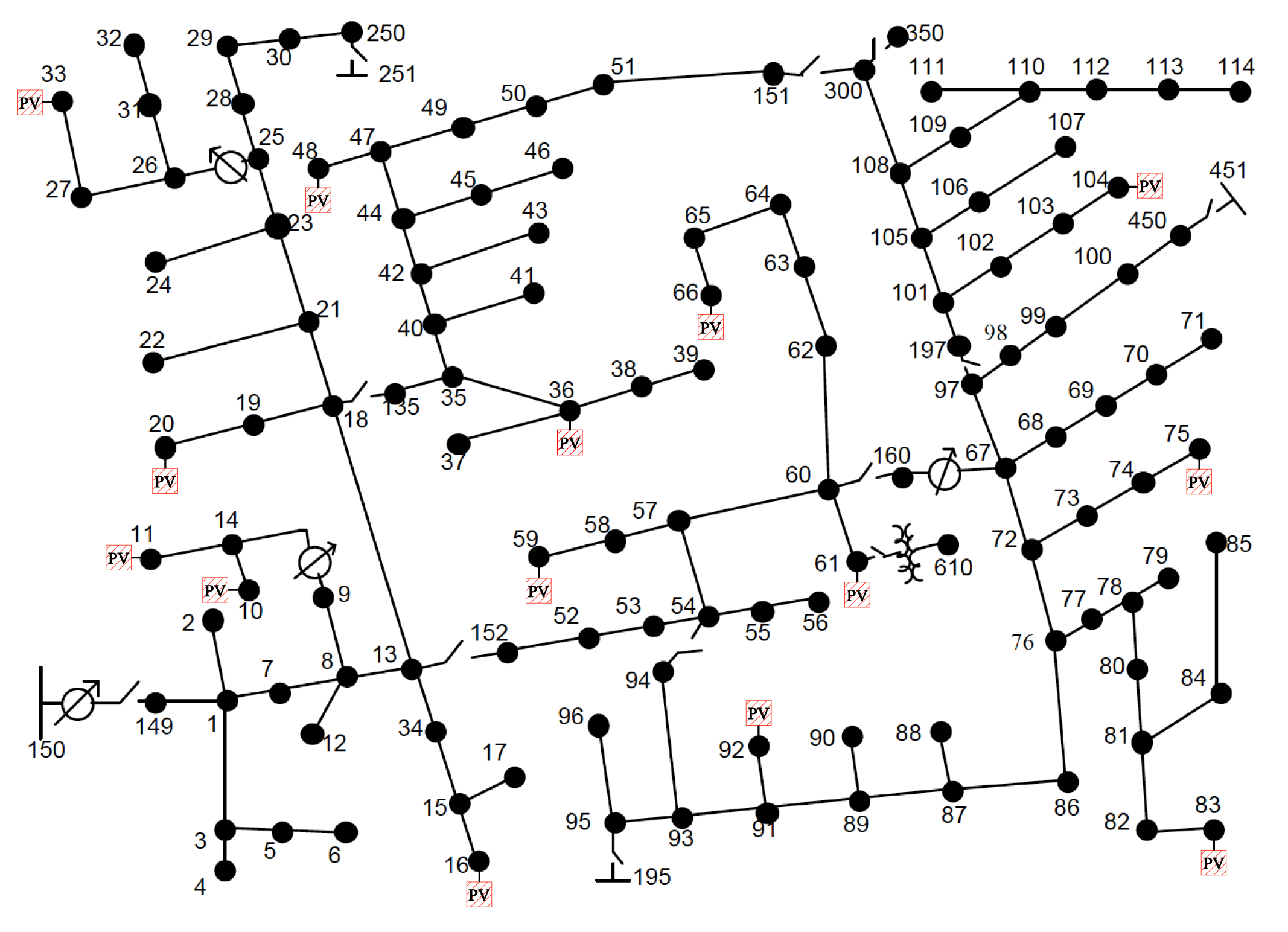}
    \caption{Schematic diagram of the IEEE 13-bus system and IEEE 123-bus system.} 
    \label{fig:13bus}
\end{figure}
\begin{figure}[h]
    \centering
    \includegraphics[width=0.4\textwidth]{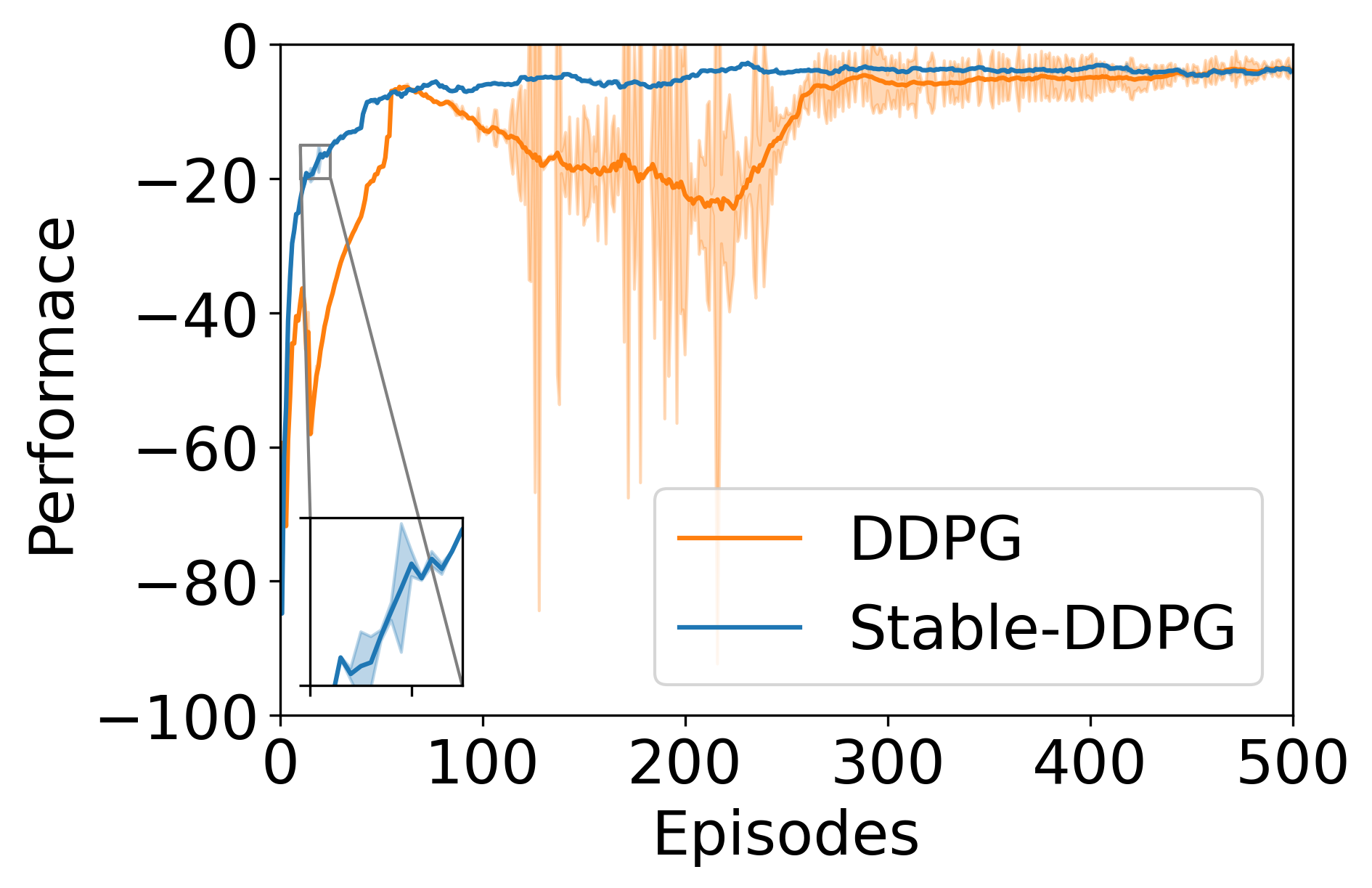}
    \caption{\revise{Model Performance vs Training Episodes. Learning process is repeated for three times. Solid line represents the mean and shaded area represents the variance.}}
    \label{fig:13bus-policy}
\end{figure}

IEEE 13-bus system is a typical radial distribution system depicted by Figure \ref{fig:13bus} (left), where three PV generators and voltage controllers are randomly picked to be located at buses 2, 7, and 9. To obtain the single-phase model, we first convert the three-phase loads to single phase loads by division, e.g. if the load has a Delta configuration, then the single-phase load is the three-phase load divided by three. For simplicity, we ignore the downstream transformer between node 3 and 6.
The nominal voltage magnitude at each bus except substation is 4.16 kV. The safe operation range $S_v$ is defined as $\pm 5\%$ of the nominal value, that is $[3.952\text{kV}, 4.368\text{kV}]$.
\revise{The overall training time for the Stable-DDPG algorithm is $71s$, and the DDPG algorithm can be trained around $85s$.} 
We first show the training curves of both learning algorithms in Figure \ref{fig:13bus-policy}. 
Stable-DDPG can quickly learn to stabilize the system with relatively low cost. It also has smaller variance compared to standard DDPG.
\begin{small}
\begin{table}[t]
\centering
\caption{Performance of linear, DDPG, and Stable-DDPG on 500 voltage violation scenarios for IEEE 13-bus system. }
 \label{table:voltage_control13b}
 \begin{tabular}{lcccc}
    \toprule
    & \multicolumn{2}{c}{Voltage recovery steps}  & \multicolumn{2}{c}{Reactive power $(\text{Mvar})$}  \\
    \cmidrule(r){2-5}
    Method     & Mean     & Std & Mean & Std \\
    \midrule
    \revise{MPC} & \revise{4.55}&\revise{8.90} &\revise{7.62} &\revise{16.40}\\
    Linear & 5.31 & 3.19 & 8.22 &  10.72 \\
    \textbf{Stable-DDPG} & \textbf{4.47} & \textbf{2.43} & \textbf{6.75} & \textbf{8.08}  \\
    DDPG &6.61  & 20.67 & 30.20  & 120.24   \\
    DDPG* &2.31 &1.18 & 3.65 & 3.21\\
    \bottomrule
\end{tabular}
Note: DDPG* denotes the performance of the DDPG policy in the subset of testing cases when it was able to stabilize the voltage.
\end{table}
\end{small}

\textbf{Model Predict Control. } We assume perfect knowledge of matrix $X$ for the IEEE 13-bus system. Considering a finite look-ahead time window $H=30$, which equals the episodic length of the RL training, the centralized Model Predict Control (MPC) algorithm can be formulated as follows.

\revise{
\small{
\begin{subequations}
\label{opt:mpc}
\begin{align}
\argmin_{\substack{\mathbf{\hat{u}}(t),\cdots, \mathbf{\hat{u}}(t+H-1),\\ \mathbf{\hat{v}}(t),\cdots, \mathbf{\hat{v}}(t+H-1)}}  &\sum_{k=0}^{H-1} \sum_{i=1}^n c_i({\hat{v}}_i(t+k), \hat{u}_i(t+k))\\
\underset{k = 0, ..., H-1}{\text{subject to}}\quad& \mathbf{\hat{q}}(t+k+1)=\mathbf{\hat{q}}(t+k)+ \Delta T \mathbf{\hat{u}}(t+k)\,, \label{eq:dynamics0}\\
    &\mathbf{\hat{v}}(t+k+1)= X \mathbf{\hat{q}}(t+k+1) + \mathbf{v}^{env}\,,  \label{eq:dynamics1}\\
    &\mathbf{\hat{v}}(t)=\mathbf{v}(t), \mathbf{\hat{q}}(t)=\mathbf{q}(t) \label{eq:initial_cond}\\
    &\underline{v}\leq \mathbf{\hat{v}}(t+H)\leq \bar{v}
\end{align}
\end{subequations}}}
\revise{
For a fair comparison, the cost function $c_i({\hat{v}}_i(t+k), \hat{u}_i(t+k))$ of MPC is chosen to be the same as the cost function used in RL training. At each time step, the finite-horizon optimal control problem \eqref{opt:mpc} is solved to obtain the control sequence. We write $\mathbf{\hat{u}}^*(t),\cdots, \mathbf{\hat{u}}^*(t+H-1), \mathbf{\hat{v}}^*(t),\cdots, \mathbf{\hat{v}}^*(t+H-1)$ as the optimal control sequence and the corresponding voltage trajectory. The control action is later selected as $\mathbf{u}(t)=\mathbf{\hat{u}}^*(t)$. We use this centralized MPC algorithm as a baseline for the IEEE 13-bus system.
}

\begin{figure}[h]
\centering
\begin{minipage}[t]{0.23\textwidth}
  \centering\raisebox{\dimexpr \topskip-\height}{%
  \includegraphics[width=\textwidth]{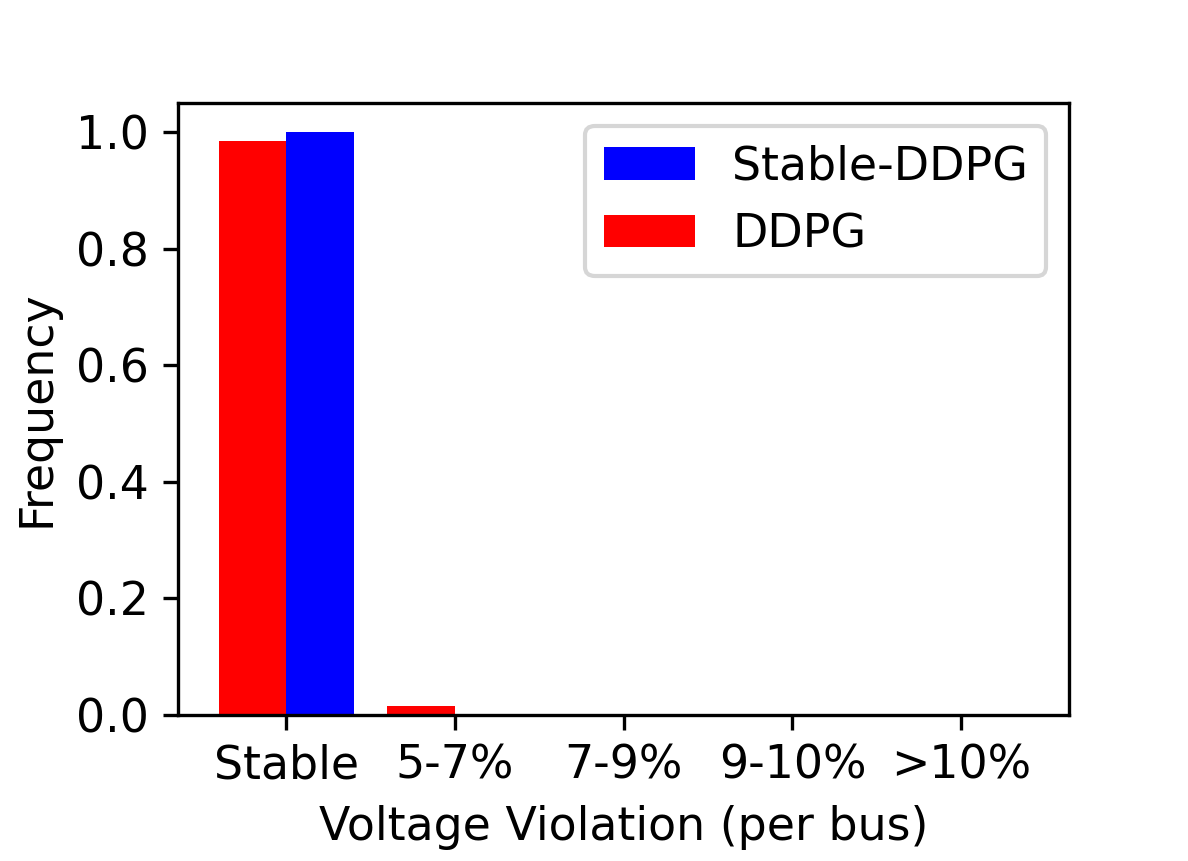}}
  \label{fig:instability}
\end{minipage}
\begin{minipage}[t]{0.23\textwidth}
  \centering\raisebox{\dimexpr \topskip-\height}{%
  \includegraphics[width=\textwidth]{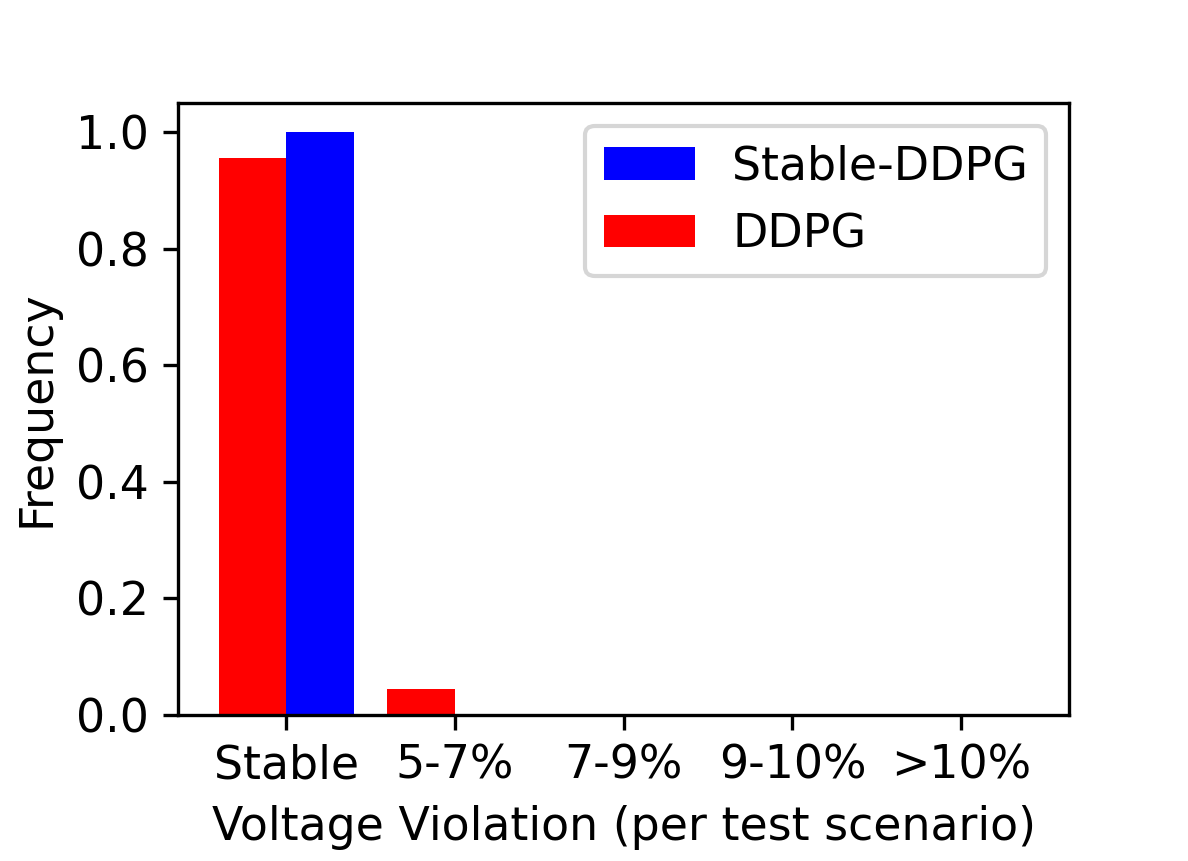}}
  \label{fig:instability}
\end{minipage}
\caption{Voltage stability for single-phase 13 bus test system. The left plot is the voltage violation for each bus, the right plot is the largest violation bus.} 
\label{fig:voltage_violation13avg}
\end{figure}

\textbf{Control Performance.} We compare the performance of the proposed Stable-DDPG method against linear policy, standard DDPG, and MPC policies on 500 different voltage violation scenarios. Table~\ref{table:voltage_control13b} shows the results. \revise{Notably, Stable-DDPG outperforms the centralized MPC algorithm even when the exact linearized system dynamics model is known for the MPC. In this case, the linearized model provides a reasonable approximation with some approximation error. 
As a result, our proposed Stable-DDPG algorithm, which interacts with the nonlinear power flow simulator for policy training can outperform the centralized MPC method. It is also worth mentioning that the computational time of the proposed Stable-DDPG (0.37ms) is on the same scale as the Linear controller (0.16ms) while significantly smaller than the MPC (449.83ms) as shown in Table \ref{tab:time}. Stable-DDPG can support a control frequency of up to 2 kHz, which enables real-time decentralized voltage control.}
Figure \ref{fig:voltage_violation13avg} demonstrates the percentage of voltage instability cases in the 500 testing scenarios. If the controller is able to bring back the voltage of all controlled buses to $[3.952\text{kV}, 4.368\text{kV}]$, the trajectory will be marked as ``stable''. Otherwise, we record the final voltage magnitudes of the controlled buses and categorized based on the violation magnitude. 
Our method achieves voltage stability in all scenarios, whereas DDPG may lead to voltage instability even in this simple setting, with the final voltage beyond the $\pm 5\%$ range for about 4\% of the test scenarios. 
\begin{table}[t]
\centering
\begin{tabular}{c|cccc}
\toprule
    \revise{Method} & \revise{MPC}&\revise{Linear} &\revise{Stable-DDPG} &\revise{DDPG}\\
    \midrule
    \revise{Time (ms)} & \revise{449.83}&\revise{0.16}&\revise{0.37}&\revise{0.17}\\
\bottomrule
\end{tabular}
\caption{\revise{Computational time comparison.}}
\label{tab:time}
\end{table}

\begin{figure}[h]
    \centering
    \includegraphics[width=0.5\textwidth]{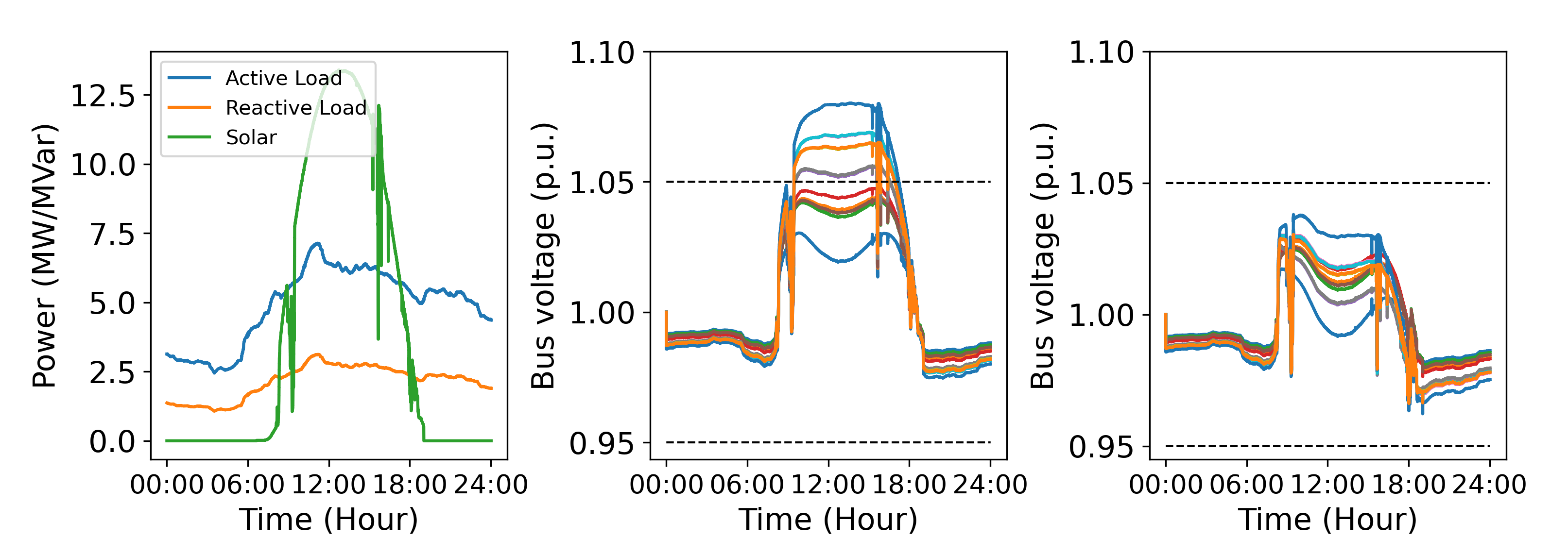}
    \caption{Stable-DDPG test with real-world load and PV dataset. The left plot is the PV and aggregated load. The right two plots are the voltage without control and with Stable-DDPG, where colored curves show voltage at different buses.}
    \label{fig:real_world}
\end{figure}

\textbf{Test with Real-world Data.} Finally, we test the proposed method using real-world data from DOE~\cite{qu2019optimal}. 
We compare the voltage dynamics without voltage control and when Stable-DDPG is used. We simulate a massive solar penetration scenario where \emph{all buses} are associated with PV and voltage controllers. The voltage control results are given in Figure~\ref{fig:real_world}. There are severe voltage violations without control, due to the high volatility in load and PV generation. In contrast, Stable-DDPG quickly brings the voltage into the stable operation range, which further demonstrates its applicability in power system voltage control. 
For the 13-bus network in Fig \ref{fig:13bus}, with a control frequency larger than $0.82$ Hz, both sides of the stability constraint $-\frac{2}{\Delta T} X^{-1}  \prec \frac{\partial \mathbf{u}}{\partial \mathbf{v}} \prec 0$ will hold.

\subsubsection{123-bus system}
We further test the controller performance in the IEEE 123-bus test feeder, which has 14 PV generators and controllers randomly selected to be placed at Buses 10, 11, 16, 20, 33, 36, 48, 59, 61, 66, 75, 83, 92, and 104. The system diagram is shown in Figure \ref{fig:13bus} (right).
The nominal voltage magnitude at each bus except substation is 4.16 kV, and the acceptable range of operation is $\pm 5\%$ of the nominal value which is $[3.952\text{kV}, 4.368\text{kV}]$. 

\begin{figure}[tb]
    \centering
    \includegraphics[width=0.48\textwidth]{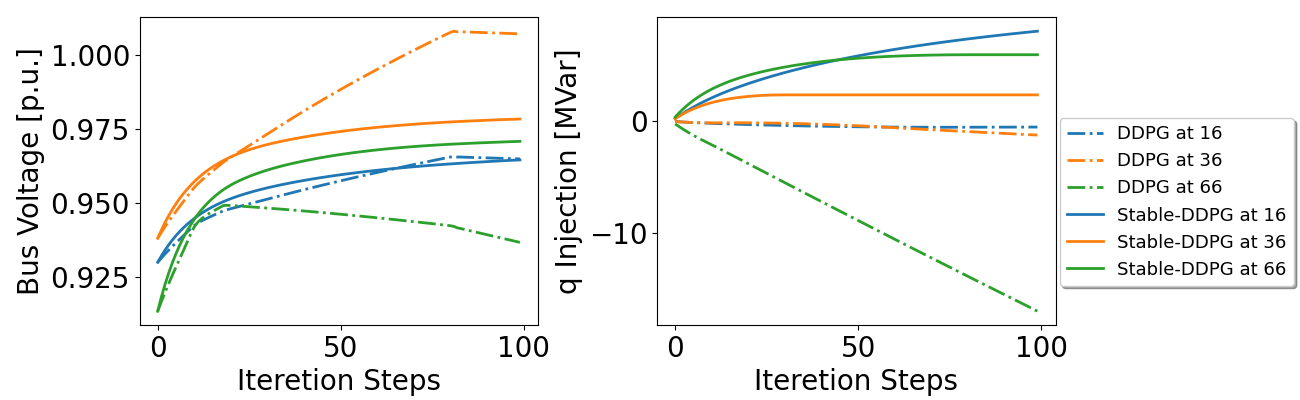}
    \caption{\revise{Stable-DDPG and DDPG were tested on a low voltage scenario simulation. The left plot is the voltage trajectories, and the right plot is the reactive power injection.}}
    \label{fig:ieee-123-low}
\end{figure}

\begin{figure}[tb]
\centering
\begin{minipage}[t]{0.23\textwidth}
  \centering\raisebox{\dimexpr \topskip-\height}{%
  \includegraphics[width=\textwidth]{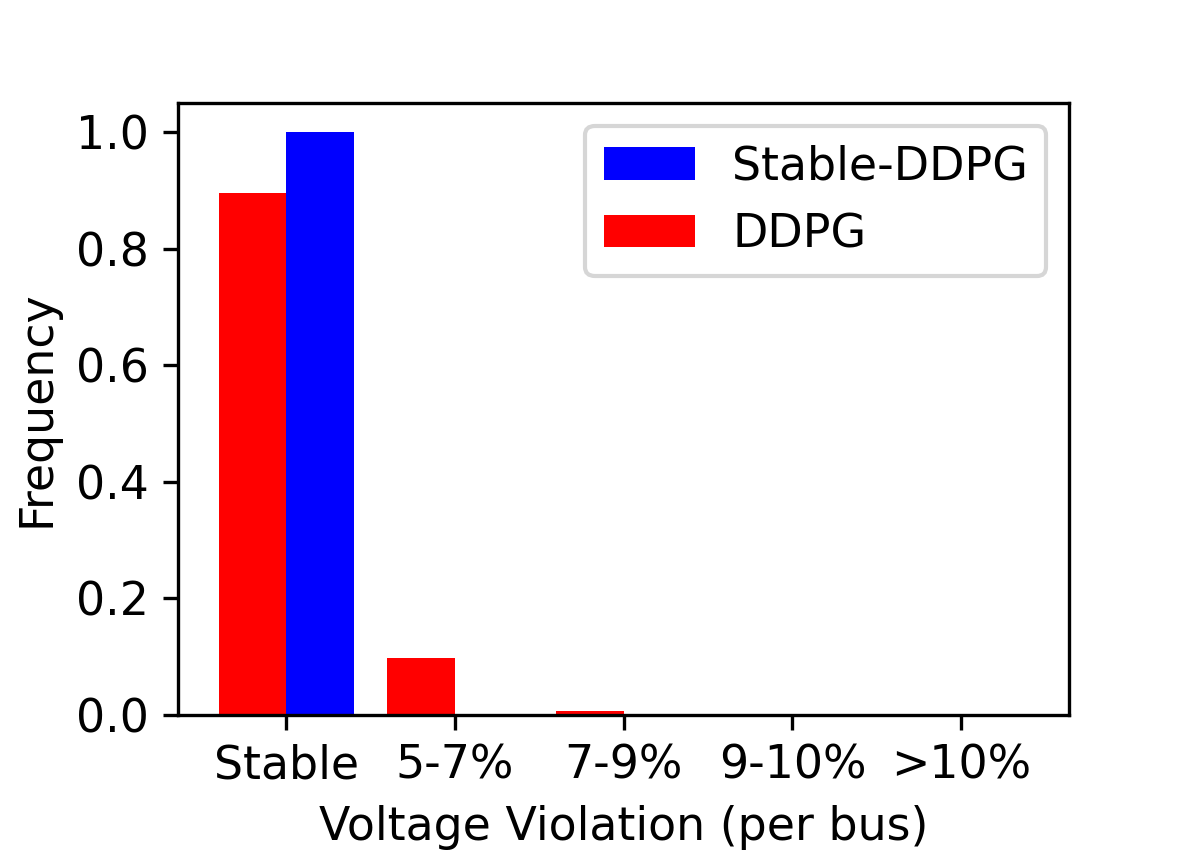}}
  \label{fig:instability}
\end{minipage}
\begin{minipage}[t]{0.23\textwidth}
  \centering\raisebox{\dimexpr \topskip-\height}{%
  \includegraphics[width=\textwidth]{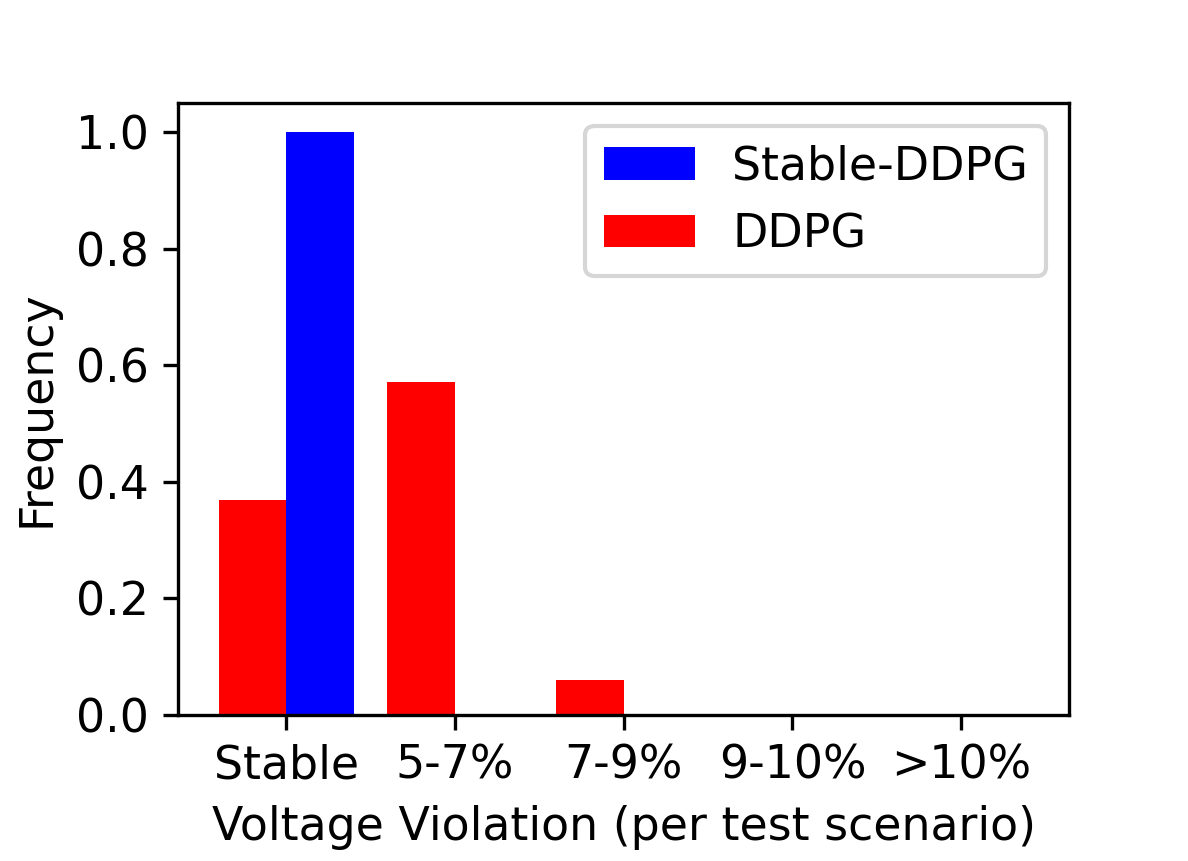}}
  \label{fig:instability}
\end{minipage}
\caption{Voltage stability for single-phase 123 bus test system.  The left plot is the voltage violation for each bus, the right plot is the largest violation bus.}
\label{fig:voltage_violation123avg}
\end{figure}

\textbf{Control Performance.} \revise{Compared with IEEE 13-bus system, the IEEE 123-bus system is more sophisticated. As a result, the computation cost for simulation is higher. 
The policy training time for the Stable-DDPG is 1450.08s  and 1300.14s for the DDPG.} Table \ref{table:voltage_control123} compares the voltage recovery time and reactive power consumption of the trained controllers. 
Although DDPG performs slightly better when it can successfully stabilize the system (denoted as DDPG*), the lack of stability guarantee can lead to oscillations and instability, thus resulting higher overall costs.
\revise{As shown in Figure \ref{fig:ieee-123-low}, the DDPG voltage controller without considering stability can lead to voltage instability, while the proposed Stable-DDPG controller shows good performance for the same test scenario.} 
\begin{table}[t]
\centering
\caption{Performance of linear, DDPG, and Stable-DDPG on 500 voltage violation scenarios with IEEE 123-bus case.}
 \label{table:voltage_control123}
 \begin{tabular}{lcccc}
    \toprule
    & \multicolumn{2}{c}{Voltage recovery steps}  & \multicolumn{2}{c}{Reactive power $(\text{Mvar})$}  \\
    \cmidrule(r){2-5}
    Method     & Mean     & Std & Mean & Std \\
    \midrule
    Linear & 41.30 & 20.30 & 1529.62 & 1302.60  \\
    \textbf{Stable-DDPG} & \textbf{32.35} & \textbf{15.40} & \textbf{1178.77} &  \textbf{992.70} \\
    DDPG & 73.91  & 36.72 &4515.33  & 2822.96   \\
    DDPG* &29.11 &22.10 & 1148.24&1357.08\\
    \bottomrule
\end{tabular}
Note: DDPG* denotes the performance of the DDPG policy in the subset of testing cases when it was able to stabilize the voltage.
\end{table}
Figure \ref{fig:voltage_violation123avg} shows that our proposed Stable-DDPG stabilizes the system voltage in all test scenarios within 100 steps. In contrast, for DDPG, about $10\%$ of buses' voltages are still beyond the $\pm 5\%$ range after a maximal control period (Fig. \ref{fig:voltage_violation123avg} Left), which accounts for approximately 63\% of test scenarios (Fig. \ref{fig:voltage_violation123avg} Right). This further highlights the necessity of explicitly considering stability in learning-based controllers. 
\subsection{Three-phase Simulation Results}
We now evaluate Stable-DDPG in three-phase systems. All simulations are built with the OpenDSS public models \cite{opendss}.
\subsubsection{13-bus system}
To stabilize all the nodes of the network, we installed a PV generator and controller in every node except the substation node. The nominal voltage magnitude and the acceptable range are the same as in the single-phase experiment. 
Table \ref{table:voltage_control} summarizes the performance of different controllers. Our proposed method achieves the best overall performance with a fast response and less reactive power consumption compared to the baseline linear policy and DDPG policy. While the DDPG algorithm has an impressive voltage recovery time and control cost if it successfully stabilizes the system (DDPG*), the percentage of stabilizing test cases is only around 34\%. 
About $16.5$\% of buses' voltages fail to recover to the nominal range that spans 66\% of 500 test scenarios, 
whereas Stable-DDPG achieves voltage stability in all scenarios. 
Furthermore, compared to the optimized linear policy, our method can save about 26.0\% in time and 35.7\% in reactive power consumption. 



\begin{table}[t]
\centering
\caption{Performance of linear, DDPG, and Stable-DDPG on 500 voltage violation scenarios with three-phase IEEE 13-bus test case.}
 \label{table:voltage_control}
 \begin{tabular}{lcccc}
    \toprule
    & \multicolumn{2}{c}{Voltage recovery steps}  & \multicolumn{2}{c}{Reactive power $(\text{Mvar})$}  \\
    \cmidrule(r){2-5}
    Method     & Mean     & Std & Mean & Std \\
    \midrule
    Linear & 19.75 & 9.10 & 46.55 & 37.76  \\
    \textbf{Stable-DDPG} & \textbf{14.61} & \textbf{3.74} & \textbf{29.94} &  \textbf{16.07}\\
    DDPG & 73.32  & 42.58 &118.44  & 74.01   \\
    DDPG* &5.39 &1.99 & 18.42 & 11.05\\
    \bottomrule
\end{tabular}
Note: DDPG* denotes the performance of the DDPG policy in the subset of testing cases when it was able to stabilize the voltage.
\end{table}

\subsubsection{\revise{123-bus system}}
Finally, we evaluate the proposed model with the unbalanced three-phase IEEE 123-bus system. The PV generator and controllers are installed in the same location as the single-phase IEEE 123-bus system. 
We summarize the control performance of different methods with the three-phase IEEE 123-bus system in Table \ref{table:voltage_control1233p}. According to the results, the average recovery time of the Stable-DDPG controller is 30\% quicker compared to the optimal linear controller. Moreover, the reactive power consumption of the Stable-DDPG is 27.7\% less than the optimal linear controller. Due to the absence of a stability guarantee, with the DDPG controller, 57.2\% of the 500 test scenarios have at least one bus that fails to recover within 100 steps, leading to a significantly longer response time and a considerable increase in reactive power consumption.
\begin{table}[t]
    \centering
    \caption{Performance of linear, DDPG, and Stable-DDPG on 500 scenarios with three-phase IEEE 123-bus test case.}
     \label{table:voltage_control1233p}
     \begin{tabular}{lcccc}
        \toprule
        & \multicolumn{2}{c}{Voltage recovery steps}  & \multicolumn{2}{c}{Reactive power $(\text{Mvar})$}  \\
        \cmidrule(r){2-5}
        Method     & Mean     & Std & Mean & Std \\
        \midrule
        Linear & 18.18 & 4.54 & 439.99 & 310.23  \\
        \textbf{Stable-DDPG} & \textbf{12.70} & \textbf{4.99} & \textbf{318.31} &  \textbf{273.22} \\
        DDPG & 59.82  & 46.46 & 4715.57  & 3993.85  \\
        DDPG* &6.12 &0.96 & 126.77&35.68\\
        \bottomrule
    \end{tabular}
    Note: DDPG* denotes the performance of the DDPG policy in the subset of testing cases when it was able to stabilize the voltage.
    \end{table}

\vspace{-6pt}
\subsection{Further Discussion}
\revise{The above results also reveal an important \emph{trade-off} between stability and the expressiveness of neural networks. DDPG algorithm with standard neural network policy obtains the best transient performance if it was able to stabilize the system (see the performance of DDPG*). However, without a stability guarantee, the DDPG controller can lead to unstable working conditions, thus incurring overall high costs compared to both optimized linear policy and Stable-DDPG  policy.} With the monotone policy network, Stable-DDPG maintains the voltage magnitude in all test scenarios at the cost of a less flexible neural network parameterization. The linear policy can be regarded as an extreme example of a restricted neural net with only one learnable parameter, its slope, and thus might get sub-optimal performance compared to the monotone neural network with more learnable parameters.

\section{Conclusion and Future Works}
In this work, we propose a stability-constrained reinforcement learning framework that formally guarantees the stability of RL 
for distribution system voltage control. The key technique that underpins the proposed approach is to use the Lyapunov stability theory and enforce the stability condition via monotone policy network design. We demonstrate the performance of the proposed method in IEEE single-phase and three-phase test systems. 
\revise{In terms of future work directions, one limitation of the proposed decentralized Stable-DDPG controller is that it can only guarantee voltage stability for the controlled buses. It is an interesting future direction to consider communications between neighboring nodes and design distributed controllers to ensure stability guarantees for buses without control. It is also a valuable future direction to unify the proposed approach in optimizing the transient cost of voltage control with steady-state cost optimization to obtain the best of both worlds. Additionally, a challenging and important task is to extend the monotone neural network design to multi-input multi-out monotone neural networks for the three-phase voltage controllers.}

\bibliographystyle{IEEEtran}
\bibliography{reference}

\vskip -2\baselineskip plus -1fil
\begin{IEEEbiography}[{\includegraphics[width=1in,height=1.25in,clip,keepaspectratio]{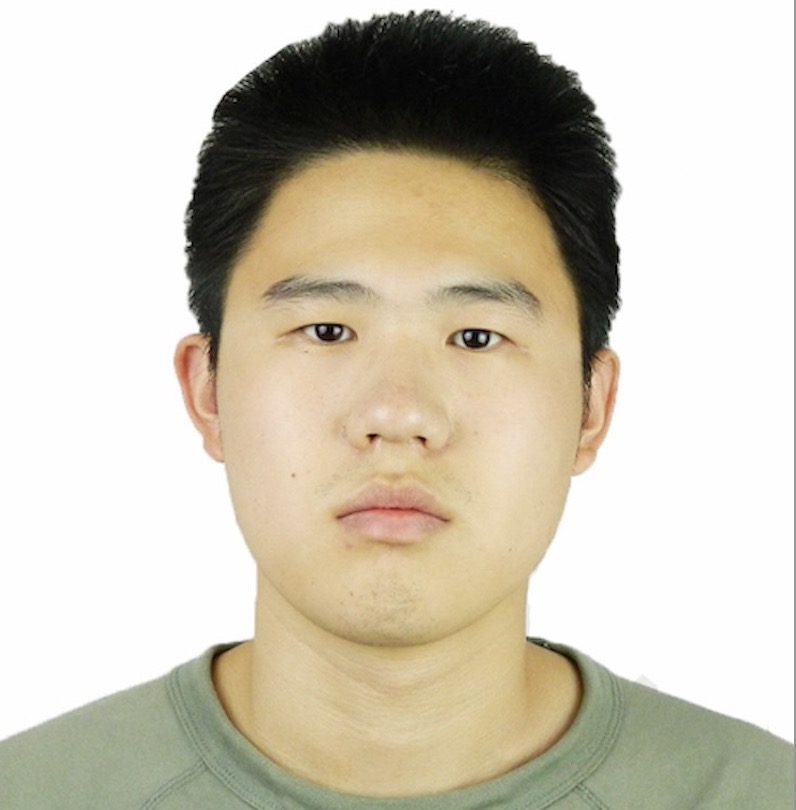}}]{Jie Feng}(Student member, IEEE) received the B.E. degree in Automation from Zhejiang University, Hangzhou, China, in 2021. He is currently pursuing his Ph.D. degree in Electrical and Computer Engineering
at the University of California, San Diego. His research interests focus on stability-constrained machine learning for power system control.
\end{IEEEbiography}

\vskip -2\baselineskip plus -1fil
\begin{IEEEbiography}[{\includegraphics[width=1in,height=1.25in,clip,keepaspectratio]{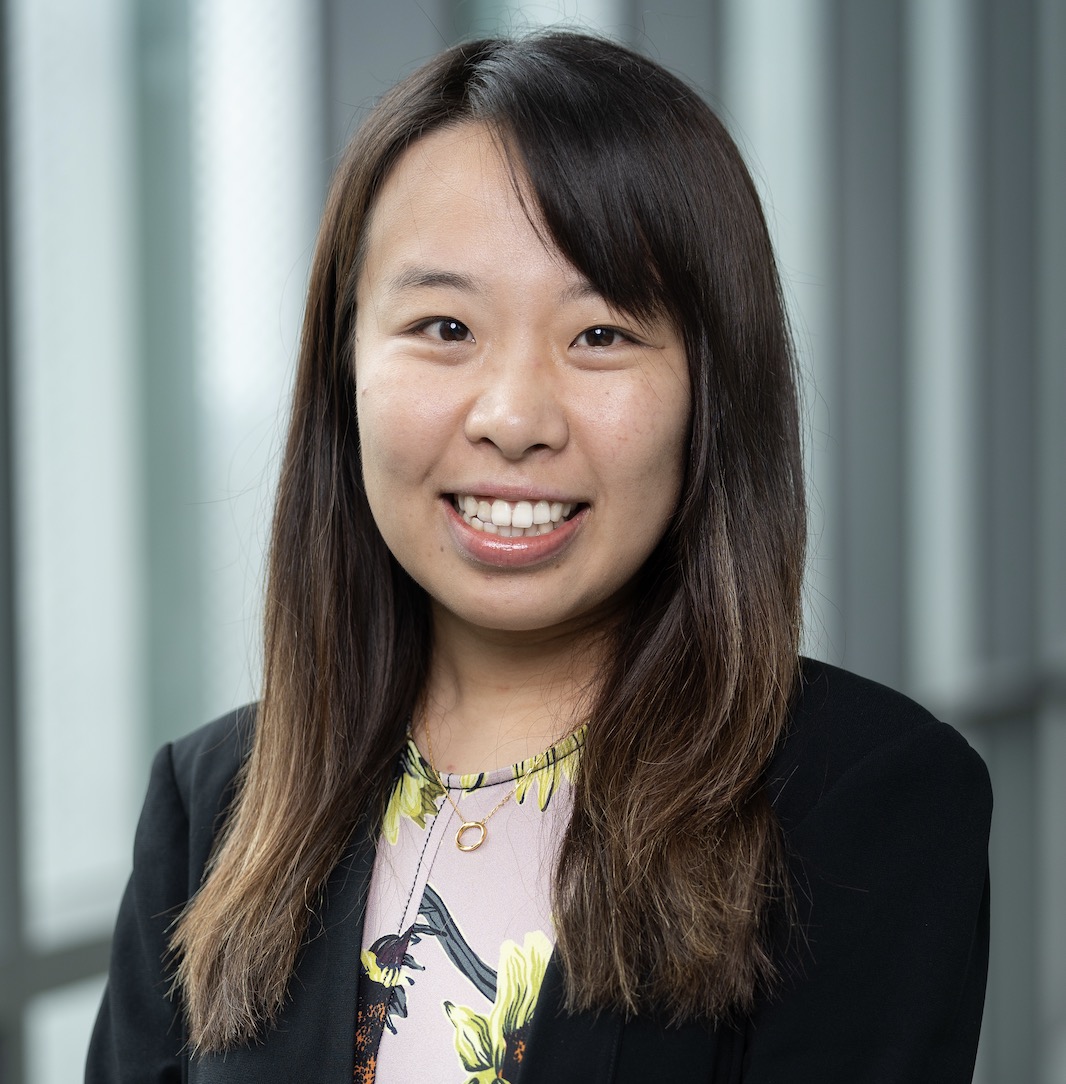}}]{Yuanyuan Shi}(Member, IEEE) is an Assistant Professor of Electrical and Computer Engineering at the
University of California, San Diego. She received her Ph.D. in Electrical Engineering, masters in Electrical Engineering and Statistics, all from the University of Washington,
in 2020. From 2020 to 2021, she was a Postdoctoral Scholar with the Department of Computing and Mathematical Sciences, Caltech. Her research interests include machine learning, dynamical systems, and control, with applications to sustainable power and energy systems.
\end{IEEEbiography}

\vskip -2\baselineskip plus -1fil
\begin{IEEEbiography}[{\includegraphics[width=1in,height=1.25in,clip,keepaspectratio]{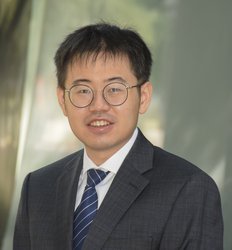}}]{Guannan Qu} (Member, IEEE) received his B.S. degree in electrical engineering from Tsinghua University, Beijing, China, in 2014, and his Ph.D. degree in applied mathematics from Harvard University, Cambridge, MA, USA, in 2019. He is currently an Assistant Professor with the Electrical and Computer Engineering Department, Carnegie Mellon University, Pittsburgh, PA, USA. From 2019 to 2021, he was a Postdoctoral Scholar with the Department of Computing and Mathematical Sciences, Caltech, Pasadena, CA, USA. His research interests include control, optimization, and machine/reinforcement learning with applications to power systems, multi-agent systems, Internet of things.
\end{IEEEbiography}

\vskip -2\baselineskip plus -1fil
\begin{IEEEbiography}[{\includegraphics[width=1in,height=1.25in,clip,keepaspectratio]{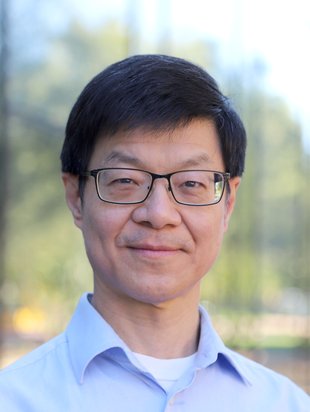}}]{Steven Low}(Fellow, IEEE) is the F. J. Gilloon Professor of the Department of Computing \& Mathematical Sciences and the Department of Electrical Engineering at Caltech.  Before that, he was with AT\&T Bell Laboratories, Murray Hill, NJ, and the University of Melbourne, Australia. He has held honorary/chaired professorship in Australia, China and Taiwan.  He was a co-recipient of IEEE best paper awards, an awardee of the IEEE INFOCOM Achievement Award and the ACM SIGMETRICS Test of Time Award, and is a Fellow of IEEE, ACM, and CSEE. He was well-known for work on Internet congestion control and semidefinite relaxation of optimal power flow problems in smart grid.  His research on networks has been accelerating more than 1TB of Internet traffic every second since 2014. His research on smart grid is providing large-scale electric vehicle charging to workplaces.  He received his B.S. from Cornell and Ph.D. from Berkeley, both in EE.
\end{IEEEbiography}

\vskip -2\baselineskip plus -1fil
\begin{IEEEbiography}[{\includegraphics[width=1in,height=1.25in,clip,keepaspectratio]{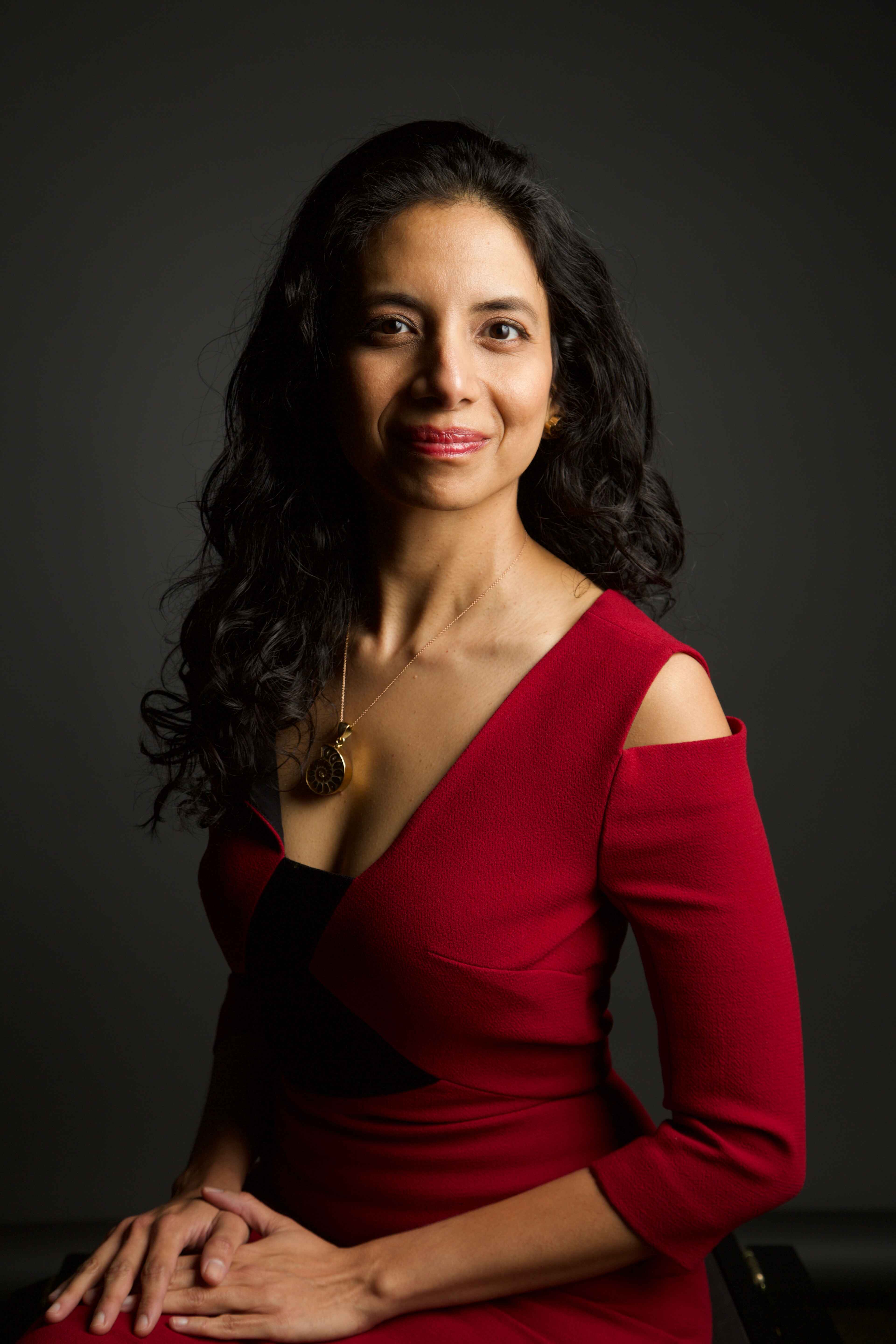}}]{Anima Anandkumar}(Fellow, IEEE) works on AI algorithms and its applications to many domains in scientific areas. She is a fellow of the IEEE and ACM, and is part of the World Economic Forum's Expert Network. She has received several awards including the Guggenheim and Alfred P. Sloan fellowships, the NSF Career award, and best paper awards at venues such as Neural Information Processing and the ACM Gordon Bell Special Prize for HPC-Based COVID-19 Research. She recently presented her work on AI+Science to the White House Science Council. 
She received her B. Tech from the Indian Institute of Technology Madras and her Ph.D. from Cornell University and did her postdoctoral research at MIT. She was principal scientist at Amazon Web Services, and is now senior director of AI research at NVIDIA, and Bren named professor at Caltech.
\end{IEEEbiography}

\vspace{-6pt}
\begin{IEEEbiography}[{\includegraphics[width=1in,height=1.25in,clip,keepaspectratio]{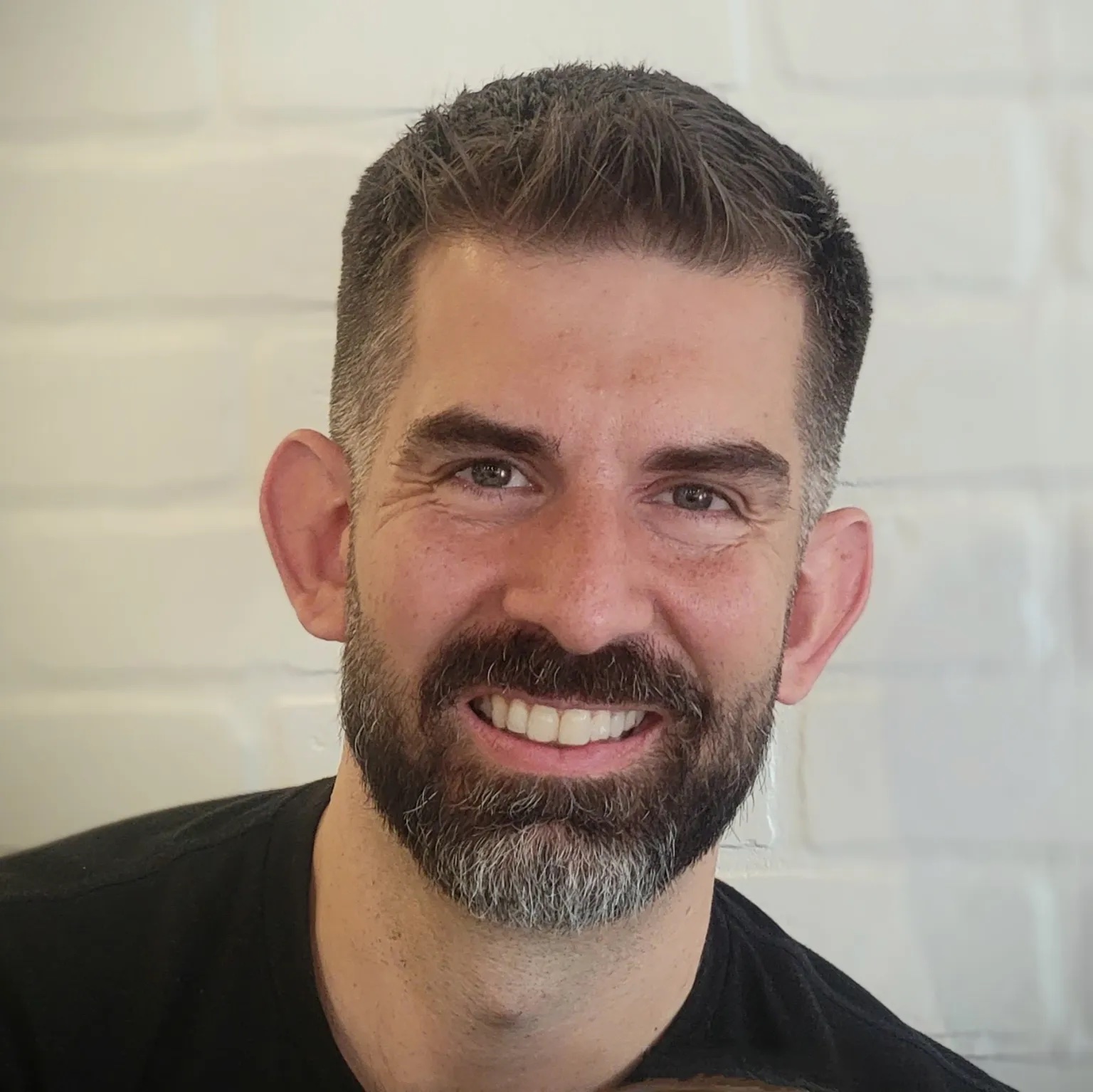}}]{Adam Wierman}(Member, IEEE) Adam Wierman is a Professor in the Department of Computing and Mathematical Sciences (CMS) at the California Institute of Technology. He is the director of the Information Science and Technology (IST) initiative and served as Executive Officer (a.k.a. Department Chair) of CMS from 2015-2020. Additionally, he serves on the Advisory Board of the Linde Institute of Economic and Management Sciences and previously served on the Advisory Board of the “Sunlight to Everything” initiative of the Resnick Institute for Sustainability. He received his Ph.D., M.Sc., and B.Sc. in Computer Science from Carnegie Mellon University in 2007, 2004, and 2001, respectively, and has been a faculty at Caltech since 2007.
\end{IEEEbiography}

\appendix


\section*{Appendix A: Proof of Theorem~\ref{thm:voltage_stab}}
\begin{proof}[Proof of Theorem~\ref{thm:voltage_stab}]
Recall the closed-loop voltage dynamics $\mathbf{v}(t+1) = f_{u}(\mathbf{v}(t))$ 
with $\mathbf{u}(t) = - g(\mathbf{v}(t))$. Let define  $h(\mathbf{v}(t))=\mathbf{v}(t)-f_u(\mathbf{v}(t))$. The Lyapunov function could be expressed compactly as follows,
\begin{equation}
     V(\mathbf{v}(t)) =  h(\mathbf{v}(t))^TX^{-1} h(\mathbf{v}(t))
\end{equation}

We further write $h(\mathbf{v}_{t+1})$\footnote{We use shorthand $\mathbf{v}_{t+1}$ instead of $\mathbf{v}(t+1)$ to simplify the notation throughout the proof.} in terms of $h(\mathbf{v}_t)$ as follows,
$$h(\mathbf{v}_{t+1})=h(\mathbf{v}_t)+\int_0^1\frac{\partial h}{\partial \mathbf{v}}(\mathbf{v}_t+t(\mathbf{v}_{t+1}-\mathbf{v}_t))(\mathbf{v}_{t+1}-\mathbf{v}_t)dt$$
From Kowalewski's Mean Value Theorem (Theorem 1 in \cite{jankovic2008mean}), we have 
$h(\mathbf{v}_{t+1})=h(\mathbf{v}_t)+J_h(\mathbf{v}_{t+1}-\mathbf{v}_t)\,,$
where $J_h=\sum_{i=1}^n\lambda_i\frac{\partial h}{\partial \mathbf{v}}(\mathbf{v}_t+k_i(\mathbf{v}_{t+1}-\mathbf{v}_t))$ for $k_i\in [0,1]$, $\lambda_i\geq 0$ for all i and $\sum_{i=1}^n\lambda_i=1$. 
Note that,
$\mathbf{v}_{t+1}-\mathbf{v}_t=f_u(\mathbf{v}_t)-\mathbf{v}_t=-h(\mathbf{v}_t)$,
Thus, we get
\begin{equation}\label{eq:time_diff_Kowalewski}
h(\mathbf{v}_{t+1}) = (I-J_h) h(\mathbf{v}_t)
\end{equation} 
Therefore,
\begin{align}
    V(\mathbf{v}_{t+1})&=h(\mathbf{v}_{t+1})^\top X^{-1} h(\mathbf{v}_{t+1}) \nonumber\\
    &=h(\mathbf{v}_t)^T(I-J_h)^TX^{-1} (I-J_h)h(\mathbf{v}_t)
\end{align}
We denote $G(\mathbf{v},\theta) = \frac{\partial f_u}{\partial v} + \frac{\partial f_u}{\partial u} \frac{\partial u}{\partial v}$ as the Jacobian of the closed-loop voltage dynamics. 
and we then define $J_G=\sum_{i=1}^n\lambda_i G(\mathbf{v}_t+k_i(\mathbf{v}_{t+1}-\mathbf{v}_t),\theta)$, where $k_i$ and $\lambda_i$ follow the definition of $J_h$. From the definition of $J_h$ and $h(\mathbf{v}_t)$, we have $J_G=I-J_h$. Thus we get
\begin{equation}
    V(\mathbf{v}_{t+1})-V(\mathbf{v}_t) = h(\mathbf{v}_t)^T(J_G^TX^{-1} J_G-X^{-1} )h(\mathbf{v}_t)
\end{equation}
With Jensen's inequality, $\forall x \in \mathbb{R}^n$, we further have 
\begin{align}
    x^TJ_G^TX^{-1}J_Gx =&\lVert X^{-1/2}J_Gx \lVert^2 \nonumber\\ 
    =&\lVert \sum_{i=1}^n\lambda_i X^{-1/2}G((\mathbf{v}_t+k_i(\mathbf{v}_{t+1}-\mathbf{v}_t),\theta)x \lVert^2 \nonumber\\
    \leq&  \sum_{i=1}^n\lambda_i \lVert X^{-1/2}G(\mathbf{v}_t+k_i(\mathbf{v}_{t+1}-\mathbf{v}_t),\theta)x \lVert^2 \nonumber\\
    =&\sum_{i=1}^n\lambda_i x^TG(\mathbf{v}_t+k_i(\mathbf{v}_{t+1}-\mathbf{v}_t),\theta)^T\backslash \nonumber\\
    &X^{-1}G(\mathbf{v}_t+k_i(\mathbf{v}_{t+1}-\mathbf{v}_t),\theta)x, 
\end{align}
Therefore, with $G(\mathbf{v},\theta)^\top X^{-1} G(\mathbf{v},\theta) - X^{-1}  \prec 0, \forall \mathbf{v} \in \mathcal{X}$, we have $V(\mathbf{v}_{t+1})-V(\mathbf{v}_t)<0$ as long as $h(\mathbf{v}_t)\neq 0$, which means the Lyapunov function is decreasing along the system trajectory. Lastly, recall that $g_{i,\theta_i}(v_i) = 0$ for $v_i \in [\underline{v}_i,\bar{v}_i]$, so $V(\mathbf{v}_{t+1})-V(\mathbf{v}_t)=0$ implies that $\mathbf{v}_t \in S_{v}$.

Given that $G(\mathbf{v},\theta)=I+ I_{\Delta T} X\frac{\partial \mathbf{u}}{\partial \mathbf{v}}$, the stability condition becomes
$$(I+ I_{\Delta T} X\frac{\partial \mathbf{u}}{\partial \mathbf{v}})^\top X^{-1} (I+ I_{\Delta T} X\frac{\partial \mathbf{u}}{\partial \mathbf{v}}) - X^{-1}  \prec 0,$$
Because of the decentralized characteristic, $\frac{\partial \mathbf{u}}{\partial \mathbf{v}}$ is a diagonal matrix. Expanding the multiplication terms, we get the stability condition as 
\begin{equation}
   -\frac{2}{\Delta T} X^{-1}  \prec \frac{\partial \mathbf{u}}{\partial \mathbf{v}} \prec 0
\end{equation}
By LaSalle's Invariance Principle and the fact that $\lim_{\mathbf{v}\rightarrow\infty}\Vert g_{\theta}(\mathbf{v})\Vert = \infty$, the stability constraint is summarized in Theorem \ref{thm:voltage_stab}.
\end{proof}

\section*{Appendix B: Experimental Details}
\begin{table}[h]
    \centering
    \begin{tabular}{cccc}
    \toprule
         Hyper-parameters& DDPG  & Stable-DDPG & Linear\\
         \midrule
         Policy network & 100-100 & 100 &1\\
         Q network & 100-100 & 100-100&100-100\\
         Discount factor $(\lambda)$&0.99&0.99&0.99\\
         Q network learning rate &  2$e^{-4}$& 2$e^{-4}$& 2$e^{-4}$\\
         Maximum replay buffer size & 1000000&1000000&1000000\\
         Target Q network update ratio &$1e^{-2}$& $1e^{-2}$& $1e^{-2}$\\
         Batch size & 256 &256&256\\
         Activation function & ReLU & ReLU & ReLU\\
    \bottomrule
    \end{tabular}
    \caption{Algorithm Hyperparameters}
    \label{tab:alg-hyper}
\end{table}
\begin{table}[htbp]
    \centering
    \begin{tabular}{cccc}
    \toprule
        Environment & Hyperparameters & DDPG & Stable-DDPG\\
        \midrule
         13bus & $\eta_1$ &1&1\\
         single-phase &$\eta_2$&100&100\\
         &Policy learning rate & 1$e^{-4}$& 1$e^{-4}$\\
         &Episode length &30&30\\
         &Training episode&500&500\\
         &State dimension $v_i$&1&1\\
         &Action dimension $u_i$&1&1\\
         \midrule
         13bus & $\eta_1$ &10&50\\
         three-phase &$\eta_2$&1000&1000\\
         &Policy learning rate & 5$e^{-5}$& 1$e^{-4}$\\
         &Episode length &30&30\\
         &Training episode&700&700\\
         &State dimension $v_i$&3&3\\
         &Action dimension $u_i$&3&3\\
         \midrule
         123bus & $\eta_1$ &0.1&0.1\\
         single-phase&$\eta_2$&100&100\\
         &Policy learning rate & 1$e^{-4}$& 1.5$e^{-4}$\\
         &Episode length &60&60\\
         &Training episode&700&700\\
         &State dimension $v_i$&1&1\\
         &Action dimension $u_i$&1&1\\
         \midrule
         \revise{123bus }& \revise{$\eta_1$ }&\revise{1}&\revise{1}\\
         three-phase&\revise{$\eta_2$}&\revise{300}&\revise{300}\\
         &\revise{Policy learning rate }&\revise{ 1$e^{-5}$}& \revise{5$e^{-5}$}\\
         &\revise{Episode length }&\revise{100}&\revise{100}\\
         &\revise{Training episode}&\revise{500}&\revise{500}\\
         &\revise{State dimension $v_i$}&\revise{3}&\revise{3}\\
         &\revise{Action dimension $u_i$}&\revise{3}&\revise{3}\\
    \bottomrule
    \end{tabular}
    \caption{Hyperparameters for Different Test Systems}
    \label{tab:env-hyper}
\end{table}
We use Pytorch to build all RL models. Table \ref{tab:alg-hyper} show the hyperparameters for of the methods. The linear policy only has one parameter which is the slope, and it is optimized with the same RL framework. The Stable-DDPG requires monotonicity of the policy network, which leads to a specially designed one-layer monotone neural network. The Q network of all three baselines and the policy network of the DDPG are designed as three-layer fully connected neural networks, the numbers of hidden units are listed in the following table. We also fine-tune hyper-parameters to obtain optimal performance of each method under different test feeders. The specific values are listed in Table \ref{tab:env-hyper}. More details about the simulation setup and model hyperparameters for all the testing cases can be found in \href{https://github.com/JieFeng-cse/Stable-DDPG-for-voltage-control}{https://github.com/JieFeng-cse/Stable-DDPG-for-voltage-control}.

%

\label{appd:A}

\end{document}